# Manifestation of strongly correlated electrons in a 2D kagome metal-organic framework

*Dhaneesh Kumar*[†, §], *Jack Hellerstedt*[†, §], *Bernard Field*[†, §], *Benjamin Lowe*[†, §], *Yuefeng Yin*[§, ±, †], *Nikhil V. Medhekar*[§, ±] *and Agustin Schiffrin*[†, §, ] *

[†]School of Physics & Astronomy, Monash University, Clayton, Victoria 3800, Australia

[§]ARC Centre of Excellence in Future Low-Energy Electronics Technologies, Monash University, Clayton, Victoria 3800, Australia

[±]Department of Materials Science and Engineering, Monash University, Clayton, Victoria 3800, Australia

* [agustin.schiffrin@monash.edu](agustin.schiffrin@monash.edu)

## Abstract

The kagome lattice, whose electronic valence band (VB) structure includes two Dirac bands and one flat band[1–3], offers a rich space to realise tuneable topological and strongly correlated electronic phases[1,4] in two-dimensional (2D) and layered materials. While strong electron correlations have been observed in inorganic kagome crystals[5,6], they remain elusive in organic systems[7,8]. The latter offer versatile synthesis protocols *via* molecular self-assembly and metal-ligand coordination[9]. Here, we report the synthesis of a 2D metal-organic framework (MOF) where di-cyano-anthracene (DCA) molecules form a kagome arrangement *via* coordination with copper (Cu) atoms on a silver surface [Ag(111)]. Low-temperature scanning tunnelling spectroscopy revealed Kondo screening – by the Ag(111) conduction electrons – of magnetic moments localised at Cu and DCA sites of the MOF. Density functional theory and mean-field Hubbard modelling show that these local moments emerge from strong interactions between kagome MOF electrons, enabling organic 2D materials as viable platforms for strongly correlated nanoelectronics and spintronics.

2D materials with a kagome crystal structure – where atoms or molecules are arranged in corner-sharing equilateral triangles (Fig. 1) – can host electronic wavefunctions that interfere destructively, resulting in highly localised electronic states[1–3]. These states, when filled gradually, give rise to increasing Coulomb electron-electron interactions, with characteristic energies that can be significantly larger than the VB width[10]. Strong Coulomb interactions between electrons populating the same localised state can lead to unpaired electron spins (i.e., local magnetic moments) and to a vast range of correlated quantum phenomena, tuneable via the VB filling, such as magnetically ordered (*e.g.*, ferro-[11], antiferromagnetic[5]) phases and metal-to-insulator Mott transitions[12].

If strong electron-electron interactions give rise to local magnetic moments in a 2D kagome material, and the latter is in contact with a metal (e.g., substrate, electrode), then the localised magnetic moments within the kagome system can potentially be screened by the metal conduction electrons via the many-body Kondo effect[5,13,14]. This effect can result in a resonance in the system's density of electronic states (DOS) at the Fermi level, which vanishes above a characteristic Kondo temperature, $T_K$. That is, the measurement of a Kondo resonance in the DOS of a 2D kagome material connected to a weakly interacting metal can reveal the presence of localised magnetic moments, and hence serve as a direct experimental probe for possible strong electron-electron interactions within the kagome system.

Recently, correlated electron phenomena and flat electronic bands have been observed in inorganic 2D kagome materials[5,6,15]. MOFs[10,16] and covalent organic frameworks[17] (COFs) with similar crystal structure and resulting electronic properties have been predicted, with the additional promise of tuneable electron-electron interactions and topology via versatile, atomically precise synthesis approaches based on metal-organic coordination[9] and organic covalent bonding[18]. While 2D kagome MOFs and COFs have been synthesised, flat bands and correlated electron phenomena have not yet been observed in these systems[8,19–21].

Here, we report the observation – via temperature-dependent scanning tunnelling spectroscopy (STS) – of Kondo screening of local magnetic moments in a 2D MOF by the conduction electrons of an underlying, weakly interacting[22] noble metal surface, Ag(111). The 2D MOF consists of a kagome arrangement of DCA molecules coordinated with Cu atoms. We show, by density functional theory (including corrections for on-site electron-electron interactions; DFT+$U$) and a mean-field Hubbard (MFH) model, that the local magnetic moments result directly from strong correlations between electrons within the kagome MOF.

We synthesised the 2D kagome MOF by depositing DCA molecules and Cu onto Ag(111) in ultrahigh vacuum (UHV; see Methods). Each DCA cyano group coordinates with a Cu atom, forming a DCA (Cu) kagome (honeycomb, respectively) arrangement with a primitive unit cell composed of 3 DCA molecules and 2 Cu adatoms[20,23,24] [see scanning tunnelling (STM) and non-contact atomic force (nc-AFM) microscopy images in Fig. 1]. We observe two types of Cu atoms (labelled $Cu_A$ and $Cu_B$ in Fig. 1) with slightly different adsorption heights [Fig. S4 in Supplementary Information (SI)]. The 2D MOF structure is not perfectly crystalline (Fig. 1a) due to incommensurability with the underlying substrate (SI Figs. S1, S2).

For insights into the local density of electronic states (LDOS) of the MOF on Ag(111), we performed differential conductance (d$I$/d$V$) STS at 4.4 K. Figure 2a shows spectra taken at the DCA centre, DCA anthracene extremity, Cu atom and MOF pore centre. The latter exhibits a broad resonance between 150 – 500 mV attributed to the confinement of the Ag(111) Shockley surface state[23] (SI Fig. S5).

In the following, we focus on d$I$/d$V$ features near the Fermi level (Fig. 2b). Spectra taken at the DCA anthracene extremities and Cu atoms show zero-bias peaks (ZBPs). Spectra at Cu$_B$ sites are qualitatively identical to those at Cu$_A$ sites but with reduced intensity; in the following we focus on Cu$_A$. At the molecule centre, we observe a pair of prominent step-like features at ~±43 mV, and satellite peaks at ~±17, ~±43 and ~±75 mV, symmetric in energy with respect to the Fermi level (with more subtle peaks at ~±56 and ~±86 mV; SI Fig. S11). These peaks are also observed at the anthracene extremity and Cu locations (grey dashed lines, Fig. 2b). Spatial mapping of these d$I$/d$V$ features is shown in Figs. 2c (zero bias) and 2d (–100 mV, below the bias onset of the step-like feature in Fig. 2b; see SI Fig. S7 for d$I$/d$V$ maps related to satellite peaks). In particular, the zero-bias d$I$/d$V$ signal at the anthracene extremities in Fig. 2c follows closely the spatial extent of the DCA lowest unoccupied molecular orbital (LUMO) observed in DCA-only monolayers on Ag(111)[22] and on Gr/Ir(111)[8].

To understand the nature of the ZBP and the energy-symmetric off-Fermi features, we acquired d$I$/d$V$ spectra at the Cu, anthracene extremity and DCA centre sites, for temperatures, $T$, between 4.4 and 145 K (Figs. 3a-c). We observed that the ZBP and off-Fermi features diminish in magnitude and broaden with increasing $T$. We fit each spectrum with the function[25]:

$$\int_{-\infty}^{\infty} \rho_{\text{fit}}(E,T) \frac{\text{d}}{\text{d}V} f_{\text{FD}}(E - eV, T) \, \text{d}E \quad (1)$$

where $f_{\text{FD}}$ is the Fermi-Dirac distribution (accounting for thermal broadening), $V$ is the bias voltage and $e$ is the absolute value of the electron charge. For the Cu and anthracene extremity sites, $\rho_{\text{fit}}$ consists of the sum of 2 pairs of Lorentzians (accounting for the energy-symmetric off-Fermi peaks) and a Fano line shape (accounting for the ZBP):

$$f_{\text{Fano}}(V) \propto \frac{(\epsilon(V) + q)^2}{\epsilon(V)^2 + 1} \quad , \quad \epsilon(V) = \frac{eV - E_0}{\Gamma_{\text{Fano}}} \quad (2)$$

where $q$ is the Fano line shape parameter, $E_0$ is the ZBP energy position, and $\Gamma_{\text{Fano}}$ is the Fano line width (SI section S6 for fitting details). For the anthracene extremity ZBP, $q = \infty$, i.e., the corresponding Fano line is a Lorentzian (Fig. 3b). For the DCA centre (Fig. 3c), $\rho_{\text{fit}}$ consists of the sum of 5 pairs of Lorentzians (energy-symmetric off-Fermi peaks) and a pair of step functions (features at ±43 mV).

The Fano line width, $\Gamma_{\text{Fano}}$, for the Cu and anthracene extremity ZBPs increases as a function of $T$ (Fig. 3d), beyond trivial thermal broadening; if the ZBP broadening were exclusively due to broadening of the Fermi-Dirac distribution, $\Gamma_{\text{Fano}}(T)$ would be constant, and the ZBP could be explained by an intrinsic MOF electronic state. We fit $\Gamma_{\text{Fano}}(T)$ with the expression:

$$\Gamma_{\text{Fano}}(T) = \sqrt{2(k_B T_K)^2 + (\pi k_B T)^2} \quad (3)$$

where $k_B$ is the Boltzmann constant and $T_K$ is a fitting parameter that we discuss below (Fig. 3d). We obtained $T_K = 125 \pm 7$ K and $139 \pm 6$ K for Cu$_A$ and the DCA anthracene extremity sites, respectively [SI Fig. S9 for Cu$_B$ site $\Gamma_{\text{Fano}}(T)$].

Equation (3) describes the temperature dependence of $\Gamma_{\text{Fano}}$ for the Fermi level resonance observed in the LDOS in the case of the Kondo effect[5,13,14]. That is, when a localised magnetic moment is screened by the spins of surrounding conduction electrons, with $T_K$ reflecting the coupling strength between the localised moment and conduction electrons' spins. Given the agreement between our temperature-dependent d$I$/d$V$ data in Fig. 3 and Eq. (3), we attribute the observed ZBP to Kondo screening of unpaired local magnetic moments within the MOF by Ag(111) conduction electrons. Our values for $T_K > 120$ K, are similar to those observed for unpaired electron spins in other π-conjugated molecule-on-metal systems[26]. The ZBP was neither observed for DCA on Ag(111) (without Cu)[22] nor in other non-kagome 2D MOFs based on Cu-cyano coordination on Ag(111)[27]; therefore, it is a property of the DCA$_3$Cu$_2$ kagome MOF.

We attribute the energy-symmetric, off-Fermi step features and peaks – observed in the d$I$/d$V$ spectra within ±100 meV of the Fermi level, prominently at the DCA centre and less so at the Kondo ZBP sites (Figs. 3a-c) – to inelastic excitation of a MOF vibrational mode and to the vibrationally-assisted Kondo effect[28], respectively, consistent with density functional theory (DFT; Methods) calculations (SI section S6 for details).

The observation of the Kondo effect in our DCA$_3$Cu$_2$/Ag(111) system provides evidence of localised magnetic moments within the MOF. We rationalised the Ag(111)-supported MOF electronic structure and these localised moments via DFT (Methods). Note that this DFT approach does not address the Kondo effect[29]. Figure 4a shows the calculated band structure of DCA$_3$Cu$_2$ on Ag(111) (without corrections for on-site Coulomb interactions between $d$ electrons, that is, with Hubbard $U$ = 0 for Cu and Ag). We calculated a transfer of 0.20 ± 0.02 electrons per primitive unit cell from MOF to Ag(111). Projection onto DCA$_3$Cu$_2$ MOF states (red circles) shows their significant contribution to the near-Fermi band structure, predominantly from the kagome-arranged DCA LUMOs (partially populated by Cu 4$s$ electrons transferred upon metal-ligand coordination), with very slight Cu 3$d$ character[16]. These calculations with $U$ = 0 yielded a negligible average local magnetic moment, $\sqrt{\langle m^2 \rangle} \approx 0.05$ $\mu_B$, per DCA molecule (Methods). Although the intrinsic kagome flat and Dirac bands are significantly perturbed by Ag(111) compared to the isolated MOF, the bandwidth defined by the MOF states (~300 meV) remains similar, with remnants of the flat band[16] still observed (in particular near the Γ-point; SI Fig. S13).

Recent theoretical work has shown electron-electron Coulomb interactions in the DCA$_3$Cu$_2$ MOF with characteristic energies of ~3 eV, several times larger than the kagome bandwidth – even at VB electron fillings where the Fermi level is far below the flat band[10]. We therefore explored the effect of electron correlations by including in our DFT calculations a Hubbard $U$ term accounting for corrections due to on-site Coulomb repulsion between electrons with Cu 3$d$ character (see Methods). Figure 4b shows the spatial distribution of magnetic moments within the DCA$_3$Cu$_2$ MOF on Ag(111) calculated for $U = 3$ eV (see Methods), closely resembling that of the DCA LUMO[22], with some Cu contribution due to partial hybridization between LUMO and $d$ orbitals. The average local magnetic moment per DCA increases monotonically as a function of $U$, reaching $\sqrt{\langle m^2 \rangle} \approx 0.28$ µ$_B$ for $U = 5$ eV (see Fig. 4c), similar to freestanding DCA$_3$Cu$_2$ (SI Fig. S17). That is, interactions between MOF electrons give rise to magnetic moments localised at DCA kagome sites. Agreement between experimental and simulated STM imaging based on these DFT+$U$ calculations (Figs. 4d, e; Methods), reinforces that the latter capture the fundamental electronic properties of the Ag(111)-supported DCA$_3$Cu$_2$ kagome MOF.

We further considered a mean-field Hubbard (MFH) model (Methods) to interpret our DFT+$U$ results, where the Hamiltonian of the freestanding 2D kagome crystal consists of a nearest-neighbour tight-binding (TB) term (hopping rate $t$ between kagome sites) and a Hubbard interaction term, $U_{MFH}$, accounting for on-site Coulomb repulsion between electrons located at kagome sites (Methods; SI section S9). Note that this is in contrast with DFT+$U$, where $U$ is a correction related to interactions between Cu 3$d$ electrons (that is, $U_{MFH} \neq U$). We used $t \approx 53.5$ meV determined by fitting the non-interacting ($U_{MFH} = 0$) TB band structure to match that given by DFT (with $U = 0$) for freestanding neutrally charged DCA$_3$Cu$_2$ (SI section S9). For $U_{MFH} > \sim 6.5t \approx 350$ meV (and electron filling matching the charge transfer given by DFT+$U$; see SI), this model shows the emergence of magnetic moments localised at kagome sites (Fig. 4f; SI section S9), with the average local magnetic moment per DCA kagome site matching closely that obtained via DFT+$U$ for DCA$_3$Cu$_2$/Ag(111) for $U \geq 3$ eV (Fig. 4c). Based on this MFH model, we found that partial disorder of the lattice (observed in our STM data; Fig. 1) does not significantly affect these local magnetic moments (SI Fig. S20).

Our DFT+$U$ and MFH calculations provide strong evidence that the local magnetic moments (revealed experimentally via the Kondo effect) result from interactions between MOF electrons. The energy scale of these interactions ($U > \sim 3$ eV in DFT+$U$, $U_{MFH} > \sim 0.35$ eV in MFH model) is larger than the kagome bandwidth, consistent with previous calculations[10]. The spatial distribution of the experimentally observed ZBPs (Figs. 2c, d) is similar to that of the DFT+$U$-derived magnetic moments (Fig. 4b), following the kagome-arranged DCA LUMOs and Cu atoms, consistent with the dominant LUMO and slight Cu 3$d$ character of the VB (Fig. 4a): on-site Coulomb repulsion between electrons *hopping* among these states give rise to localised magnetic moments that become Kondo-screened by Ag(111) conduction electrons (Fig. 5). Our calculations indicate a local magnetic moment of ~1/4 $\mu_B$ per DCA (Fig. 4c); Kondo screening has been demonstrated for fractional magnetic moments of similar magnitude, experimentally[30] and theoretically[31].

Our DFT+$U$ calculations indicate (frustrated) antiferromagnetic coupling between MOF local magnetic moments (Fig. 4b), with an estimated characteristic energy scale < 1 meV (SI section S11). We also estimated that the energy scale of any substrate-mediated Ruderman-Kittel-Kasuya-Yosida interspin coupling is < 10 meV. These energy scales are significantly smaller than the estimated Kondo exchange interaction (~250 meV; SI section S11); Kondo screening is the dominant effect and quenches magnetic order. Given these energy scales, we also rule out that the off-Fermi energy-symmetric d$I$/d$V$ features in Figs. 3a-c are related to inelastic spin (*e.g.*, triplet-to-singlet) excitations[14].

Our DCA$_3$Cu$_2$/Ag(111) system differs from DCA$_3$Cu$_2$ on Cu(111)[23,24] (SI Figs. S1, S2) and other similar systems[8,20], where the Kondo effect and strong electron-electron interactions were not observed. Compared to Cu(111), the kagome MOF interacts and hybridizes less with Ag(111), consistent with the difference in chemical reactivity of both surfaces[32] and previous studies of DCA/Ag(111)[22] and DCA/Cu(111)[23]. The balanced interactions with Ag(111) allow the MOF to retain its fundamental electronic properties, with Ag(111) providing, simultaneously, a reservoir of conduction electrons that enables the Kondo effect. The latter serves as a direct experimental probe of local magnetic moments which, as shown by both our DFT+$U$ and MFH calculations, emerge from strong electron correlations within the 2D kagome MOF. Our work represents a significant step towards tuneable electron correlations – and hence controllable electronic and magnetic quantum phases – in self-assembled 2D organic materials.

Figures

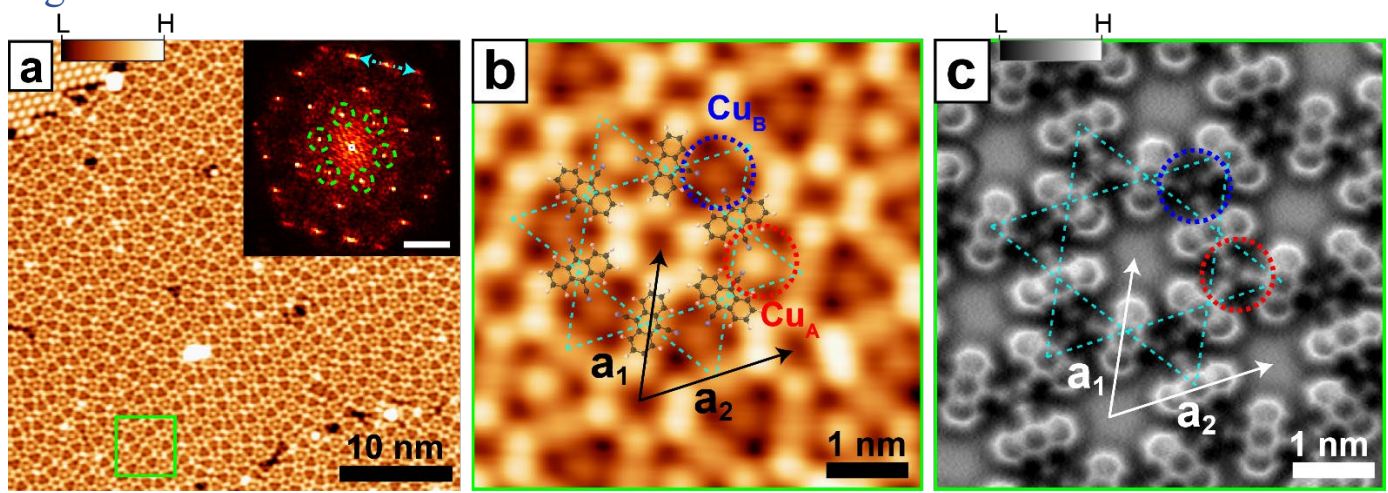

**Figure 1. Atomically thin kagome metal-organic framework (MOF) on a noble metal surface. a**, STM image of the DCA$_3$Cu$_2$ kagome structure on Ag(111) ($V_b = -20$ mV, $I_t = 50$ pA). Inset: Fourier transform (FT; $|\mathbf{k}|$ = 1/wavelength; scale bar: 1 nm$^{-1}$) of STM image. Green dashed circles indicate 1$^{st}$ harmonic peaks corresponding to the MOF hexagonal lattice. Anisotropic stretching of peaks (dashed cyan arrow) and diffuse background (around $\mathbf{k} = 0$) of FT are due to lack of perfect long-range order. **b – c,** STM (b; $V_b = -20$ mV, $I_t = 50$ pA) and tip-functionalised (CO) nc-AFM (c; tip retracted 0.3 Å with respect to STM set point, $V_b = 3$ mV, $I_t = 150$ pA, adjusted on top of DCA molecule centre) images of region indicated by green box in (a), with DCA chemical structure superimposed (black: carbon; blue: nitrogen; white: hydrogen). Vectors $\mathbf{a_1}$, $\mathbf{a_2}$ define a primitive unit cell ($\|\mathbf{a_1}\| = \|\mathbf{a_2}\| = 2.045$ nm; $\sphericalangle(\mathbf{a_1}, \mathbf{a_2}) = 60°$). Dashed cyan lines show the molecular kagome arrangement. Cu atoms can have two different STM and nc-AFM apparent heights differing by ~19 pm ('Cu$_A$' and 'Cu$_B$', dashed circles; SI Fig. S4).

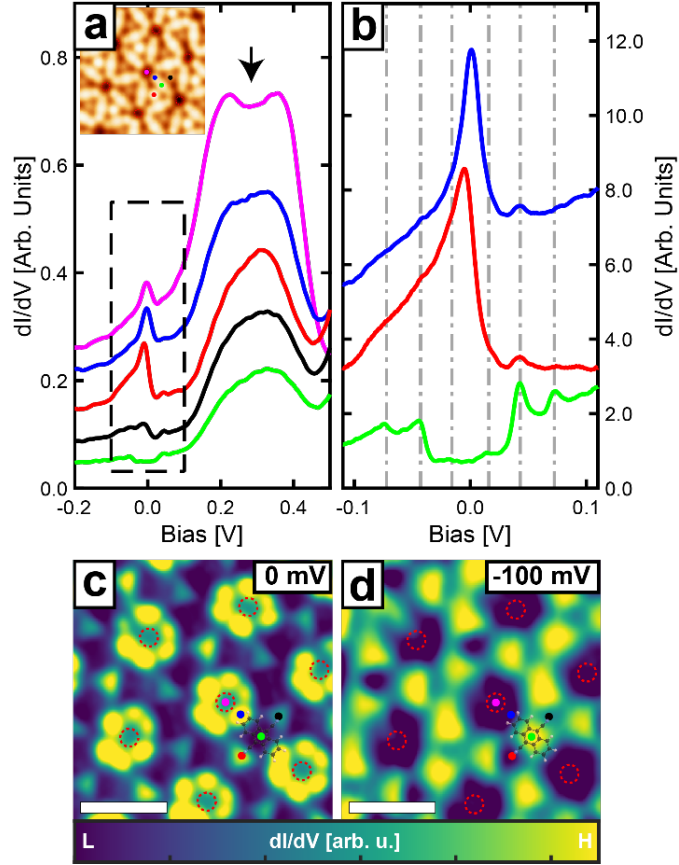

**Figure 2. Electronic properties of DCA$_3$Cu$_2$ kagome MOF on Ag(111): local density of states resonances near the Fermi level. a**, d$I$/d$V$ STS spectra (set point $V_b$ = –250 mV, $I_t$ = 0.5 nA; offset for clarity) at high symmetry points of kagome MOF (indicated in inset STM image; $V_b$ = –200 mV, $I_t$ = 0.5 nA). Broad resonance between 0.15 – 0.5 V (black arrow) attributed to Ag(111) Shockley surface state confined within kagome pores (see SI section S2). **b**, Near-Fermi d$I$/d$V$ STS spectra [dashed rectangle in (a)] at DCA anthracene extremity (blue) and MOF Cu (Cu$_A$) atom (red), showing sharp zero-bias resonances, and at DCA centre (green), showing a prominent step at ~±43 mV and peaks at ~±17, ±43 and ±75 mV (indicated by vertical grey dashed lines, and related to inelastic tunnelling due to DCA vibrational modes; see SI section S6). **c – d,** d$I$/d$V$ maps at 0 and –100 mV (following STM topography with tip retracted 120 and 30 pm, respectively, with respect to set point $V_b$ = –200 mV, $I_t$ = 0.5 nA; see Methods).

At 0 mV (–100 mV), d$I$/d$V$ intensity is higher at Cu atoms and anthracene extremities (anthracene centres, respectively). Red dashed circles: pores of kagome MOF. Scale bars: 1.5 nm.

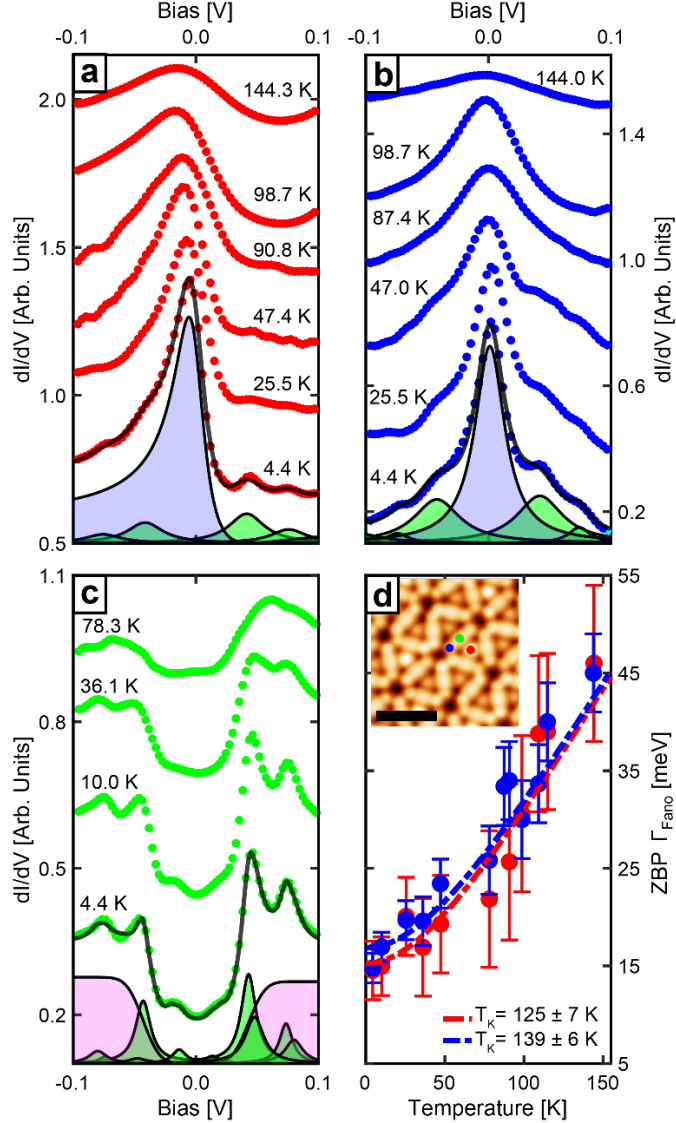

**Figure 3. Temperature-dependent near-Fermi d$I$/d$V$ spectra: evidence of the Kondo effect. a – c,** Background-subtracted d$I$/d$V$ spectra at Cu (red data points; here, Cu$_A$), DCA anthracene extremity (blue) and DCA centre (green), taken at various temperatures, $T$. Black curves: total fits of $T$ = 4.4 K data, consisting of a Fano line shape for the ZBP (blue filled curves), pairs of Lorentzians for the energy-symmetric off-Fermi peaks (green filled) and a pair of error functions for the energy-symmetric step functions (magenta filled). **d,** Width, $\Gamma_{Fano}$, of ZBP fit Fano line shape in (a) and (b), for Cu (red) and DCA anthracene extremity (blue), as a function of $T$. Effect of thermal broadening is accounted for (see SI section S6). Dashed curves: fitting of $\Gamma_{Fano}(T)$ with Eq. (3) in text, resulting in $T_K = 125 \pm 7$ K and 139 $\pm$

6 K for $Cu_A$ and anthracene extremity, respectively. Inset: STM image of $DCA_3Cu_2$ MOF ($V_b = -250$ mV, $I_t = 1.5$ nA), with locations at which d$I$/d$V$ spectra in (a) – (c) were obtained; scale bar: 2 nm.

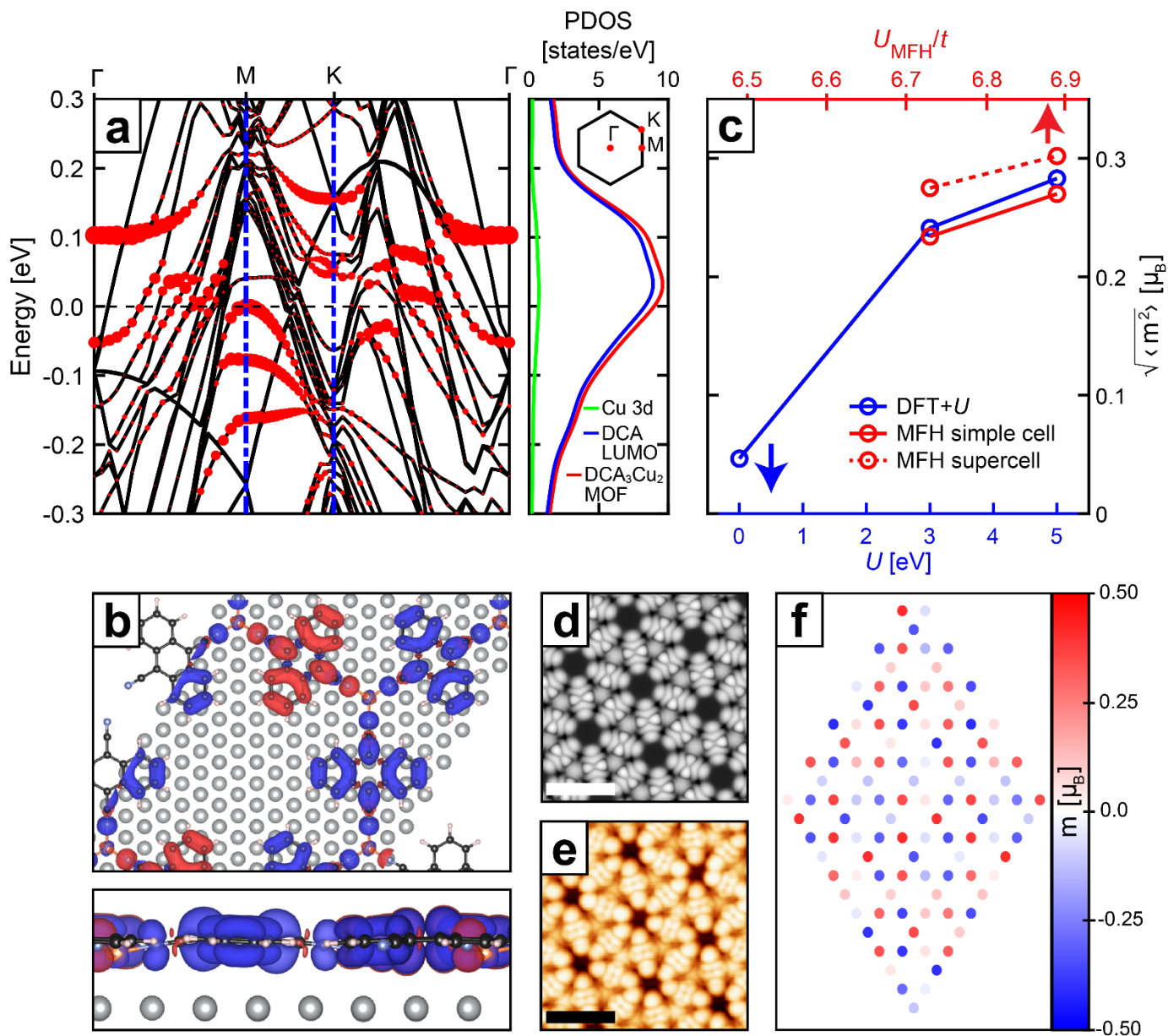

**Figure 4. Theoretical calculations: localised magnetic moments due to strong electron-electron interactions. a,** Band structure (along reciprocal space high-symmetry points) and projected density of states (PDOS) of $DCA_3Cu_2$ MOF on Ag(111) calculated by DFT ($U = 0$). Red circles and curve correspond to projections onto $DCA_3Cu_2$ MOF orbitals. Kagome bands have predominantly DCA LUMO character, with slight Cu $3d$ contribution due to partial hybridisation between LUMO and Cu $3d$ orbitals. **b,** Ground state magnetic moment density isosurface ($2\times10^{-4}$ $\mu_B/\text{Å}^3$; top and side views) of $DCA_3Cu_2$/Ag(111)

calculated *via* DFT+$U$ ($U$ = 3 eV). Note magnetic moment orientations (red: up; blue: down) indicating frustrated antiferromagnetic coupling. **c**, Average magnetic moment per DCA molecule, calculated for DCA$_3$Cu$_2$/Ag(111) *via* DFT+$U$ as a function of $U$ (blue; single unit cell; see Methods), and *via* mean-field Hubbard (MFH) model [solid red curve: primitive unit cell consisting of 3 kagome sites; dashed red curve: large supercell as in (f)] for freestanding DCA$_3$Cu$_2$ with $U_{MFH}$ = 6.73$t$ and 6.89$t$ (electron filling of 0.3, tuned to match MOF-to-surface electron transfer of ~0.20e$^-$ per primitive unit cell given by DFT+$U$ for DCA$_3$Cu$_2$/Ag(111) with $U$ = 3 and 5 eV; see Methods and SI). The larger average local magnetic moment for the supercell results from the greater magnetic configurational freedom (SI section S9). **d**, Simulated STM image of DCA$_3$Cu$_2$/Ag(111) derived from DFT+$U$ calculations ($U$ = 3 eV; $V_b$ = –100 mV; see Methods). **e**, Experimental STM image of DCA$_3$Cu$_2$/Ag(111) (tip functionalised with a CO molecule, $V_b$ = –20 mV, $I_t$ = 25 pA). Scale bars: 2 nm. **f**, Ground state magnetic moments for DCA kagome sites in freestanding DCA$_3$Cu$_2$ MOF (supercell), calculated by MFH model ($U_{MFH}$ = 6.73$t$, electron filling: 0.30).

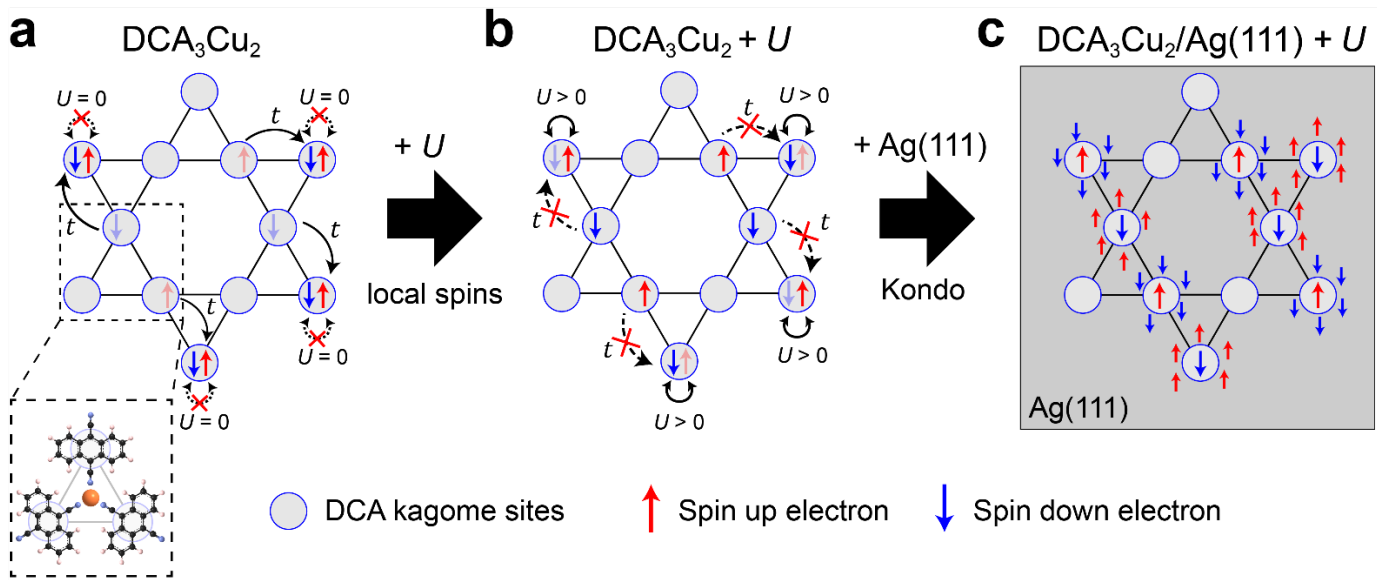

**Figure 5. Emergence of local magnetic moments from electron-electron interactions and Kondo screening in kagome DCA₃Cu₂ MOF on Ag(111). a**, Schematic of 2D kagome lattice, where each circle represents a DCA site within the freestanding DCA₃Cu₂ MOF (dashed box inset: correspondence with DCA₃Cu₂ structure). Absence of on-site electron-electron Coulomb repulsion ($U = 0$) facilitates electron hopping (rate $t$), allowing for two electrons to occupy the same kagome site, resulting in a zero net magnetic moment ($m = 0$; arrows represent electron spin). **b**, Nonzero electron-electron interactions ($U > 0$) hinders inter-site electron hopping, resulting in non-zero magnetic moments ($m \neq 0$) localised at kagome sites. **c**, On Ag(111), the spin of surface conduction electrons (represented by arrows surrounding circles) can screen the local magnetic moments within the kagome MOF, resulting in a Fermi-level LDOS Kondo resonance observed *via* d$I$/d$V$ STS.

## Methods

**Sample Preparation**

The atomically thin $DCA_3Cu_2$ kagome MOF on Ag(111) was synthesised in ultrahigh vacuum (UHV) by depositing DCA molecules (Tokyo Chemical Industry; > 95% purity) and Cu atoms (>99.99% purity) from the gas phase onto a clean Ag(111) surface (obtained by repeated cycles of $Ar^+$ sputtering and annealing at 790 K). We considered two different sample preparation procedures: (i) co-deposition of DCA and Cu onto Ag(111) held at room temperature; (ii) sequential deposition of DCA followed by Cu onto Ag(111) held at ~100 K. Procedure (i) yielded < 10% MOF coverage, with high coverage of DCA-only domains. Procedure (ii) yielded > 60% MOF coverage. Subsequent annealing at RT led to significantly less MOF coverage and an increased amount of Cu clusters and DCA-only domains, indicating that the kagome $DCA_3Cu_2$ structure on Ag(111) is metastable, as previously suggested[33]. Base pressure during molecular and metal depositions was < $5 \times 10^{-10}$ mbar.

**STM & d$I$/d$V$ STS Measurements**

All STM and d$I$/d$V$ STS measurements were performed at 4.4 K in UHV (< $1 \times 10^{-10}$ mbar) with an Ag-terminated Pt/Ir tip, unless otherwise stated. Topographic STM images were acquired in constant-current mode with sample bias reported throughout the text. All d$I$/d$V$ STS measurements, unless otherwise stated, were performed by recording the tunnelling current as a function of tip-sample bias voltage in the junction, with tip-sample distance stabilised with respect to a specified tunnelling current and bias voltage set point at a fixed reference location. We then numerically differentiated the $I$-$V$ data to obtain d$I$/d$V$ as a function of bias voltage. All d$I$/d$V$ maps were obtained with a lock-in technique by applying a small modulation of 10 mV (peak amplitude) at 1.13 kHz frequency to the bias voltage. The d$I$/d$V$ maps in Fig. 2 were acquired using a multi-pass (MP) approach. This technique consists of: (i) acquiring a constant-current STM topographic profile along a scanning line (set point $V_b = -200$ mV, $I_t = 500$ pA, scanning speed 1.6 nm s$^{-1}$);

(ii) recording d$I$/d$V$ with lock-in amplifier (10 mV bias voltage modulation, 1.13 kHz; see above) while scanning (speed: 0.8 nm s$^{-1}$) the same line and following the same constant-current STM topographic profile as in (i) but with the tip retracted 120 pm and 30 pm for Figs. 2c and 2d, respectively; and (iii) repeating this procedure sequentially for each scanned line of the map. This approach has the benefit of minimising variations of d$I$/d$V$ due to changes in STM apparent topography. The STM image in Fig. 4e was obtained using a tip functionalised with a carbon monoxide (CO) molecule (see below).

### Nc-AFM Measurements

All nc-AFM measurements were performed at 4.4 K in UHV (< 1 x 10$^{-10}$ mbar) with a qPlus sensor (resonance frequency, $f_0$ ~ 30 kHz, spring constant, $K$ ~ 1.8 kN m$^{-1}$, amplitude modulation: 60 pm) with an Ag-terminated Pt/Ir tip. Unless otherwise stated, we functionalised the tip with a CO molecule[34] by introducing CO gas into the UHV chamber (5 x 10$^{-8}$ mbar for ~3 s) with the sample at 4.4 K and picking up a CO molecule on bare Ag(111) [or bare Cu(111); see SI] with the tip. Nc-AFM imaging was performed at a constant tip height stabilised with respect to a specified tunnelling current and bias voltage set point at a fixed reference location.

### Temperature-Dependent d$I$/d$V$ STS

The base temperature of our system was kept at 4.4 K for d$I$/d$V$ STS measurements performed for temperatures up to ~70 K. The STM was heated to a specified temperature (by applying power to built-in Zener diode) and was stabilised at that temperature for 1 – 1.5 hours to allow thermal drift to settle prior to data acquisition. At higher temperatures, we employed a drift correction software to compensate for any residual thermal drift. For STS measurements performed above 77 K, the STM was heated from a base temperature of 77 K. For all temperature-dependent d$I$/d$V$ STS data, the same STM set point of $V_b$ = –250 mV, $I_t$ = 1.5 nA was used. Effects of thermal broadening were included in the fitting of d$I$/d$V$ spectra[25] (see main text and SI section S6).

## Density Functional Theory (DFT) Calculations

We performed density functional theory (DFT) calculations as implemented in the Vienna Ab-initio Simulation Package (VASP)[35], with the Perdew-Burke-Ernzerhof (PBE) functional under the generalised gradient approximation (GGA) to describe exchange-correlation effects[36] and projector augmented wave (PAW) pseudopotentials[37,38]. A 400 eV cut-off was used for the plane wave basis set. A semi-empirical potential developed by Grimme (DFT-D3) was used to describe van der Waals forces[39]. For an enhanced description of correlations between $d$ electrons, we used Dudarev's implementation[40] of DFT+$U$. We considered a systematic variation of $U$ (see Fig. 4c) which represents a correction energy term accounting for interactions between Cu $3d$ electrons (we used $U = 0$ for the substrate unless otherwise indicated; see SI). Values of $U \geq 3$ eV for Cu $3d$ are consistent with literature[41].

Structural relaxations were performed with a 3×3×1 Monkhorst-Pack Γ-centred $k$-points grid, with ionic positions relaxed until the Hellman-Feynman force on each atom was < 0.01 eV/Å. Relaxation of freestanding $DCA_3Cu_2$ ($DCA_3Cu_2$ on metal substrate) used Gaussian smearing with a width of 0.05 eV (first order Methfessel Paxton smearing with a width of 0.2 eV, respectively).

For calculations on a metal substrate, we used a three-atom-thick metal slab for the substrate, with hydrogen atoms on the bottom surface as a passivation layer. Only the two top layers were structurally relaxed; the bottom layer of metal atoms were kept fixed at their bulk coordinates. We included 15 Å of vacuum spacing above the slab to minimise possible interactions with its image resulting from periodic boundary conditions.

For $DCA_3Cu_2$ kagome on Ag(111), a supercell of 7×7 Ag surface atoms was used, with the experimental bulk lattice constant[42] of 2.889 Å, and with the MOF commensurate with the underlying substrate (periodic boundary conditions required by computation; see Fig. 4b). For $DCA_3Cu_2$ kagome on Cu(111), a supercell of 8×8 Cu surface atoms was used, with the experimental bulk lattice constant[42] of 2.554 Å, and a MOF adsorption geometry given by our nc-AFM measurements (SI Fig. S2b).

We calculated the electronic structure of $DCA_3Cu_2$ [in the freestanding case, on Ag(111), and on Cu(111); without spin polarisation] using a grid of 7×7×1 $k$-points, self-consistently until the energy changed by less than $10^{-4}$ eV between steps. Tetrahedral interpolation with Blöchl corrections was used for Brillouin zone integration. These parameters allowed for convergence of charge density. For calculations with spin polarisation, an 11×11×1 $k$-point grid and an energy convergence criterion of $< 1\times10^{-6}$ eV were used due to small energy scales involved for different spin configurations. The DOS was calculated with a 15×15×1 $k$-point grid and Gaussian smearing with a width of 0.05 eV while keeping the charge density constant. Dipole corrections to the electrostatic potential perpendicular to the slab were included in the electronic structure calculations.

For calculations of spin density of the $DCA_3Cu_2$ MOF (e.g., Fig. 4b), we considered initial conditions where each DCA molecule, in a single unit cell, was either positively spin polarised (1), negatively spin polarised (–1) or had zero spin polarisation (0). This was achieved by setting the magnetic moment for each carbon atom in a single DCA molecule to $+1$ $\mu_B$, $-1$ $\mu_B$ or $0$ $\mu_B$, respectively. We trialled each unique combination of initial DCA spin polarisation configuration in a single unit cell up to symmetry [i.e., (0, 0, 0), (1, 0, 0), (1, –1, 0), (1, 1, 0), (1, 1, –1), (1, 1, 1); 3 DCA molecules in a single unit cell]. We recorded the lowest energy magnetic configuration after calculating to self-consistency.

Charge and magnetic moments were calculated by partitioning using DDEC6 in order to assign spin and charge to the components of the $DCA_3Cu_2$ kagome MOF and the metal substrate[43–46]. We also performed partitioning with Bader charge analysis for comparison[47]. We found that the electron transferred from MOF to metal surface [$0.20 \pm 0.02$ electrons per primitive unit for $DCA_3Cu_2$/Ag(111); SI section S8], calculated by DDEC was more robust than that calculated by Bader. The calculated electron transfer varied, however, slightly as a function of $U$ and is reflected in the uncertainty of the electron transfer. Spin analysis was consistent between Bader and DDEC. The magnetic moment on each DCA molecule was calculated as the sum of magnetic moments on the atoms of that DCA molecule. The average local magnetic moment, $\sqrt{\langle m^2 \rangle}$, was the root-mean-square of all the DCA magnetic moments in a single unit cell.

For isolated molecules, such as DCA (or Cu$_2$DCA in SI), calculations were performed by only considering the $\Gamma$ point, with a Gaussian smearing width of 0.05 eV and no van der Waals corrections. Vibrational modes (SI) were calculated via the Hessian matrix and central differences with a step size of 0.015 Å, as implemented in VASP.

All structure visualization was performed with the VESTA software package[48]. PyProcar was used for post-processing of projected band structures[49].

**Mean-Field Hubbard Model**

To gain insight into how electron-electron interactions, unit cell size (e.g., primitive or supercell), electron filling, and disorder affect the electronic and magnetic properties of the kagome MOF, we used a Hubbard model which we solved via the Hartree-Fock mean-field approximation assuming collinear spins (equivalent to Hartree theory). This approximation assumes a single-electron problem with a potential depending on electron density, determined self-consistently from the single-electron eigenstates[50]. The mean-field Hubbard (MFH) model treats many-body effects similarly to DFT, albeit with a simpler potential. The MFH Hamiltonian is:

$$H = -t \sum_{\langle i,j \rangle, \sigma} \left( c_{i,\sigma}^{\dagger} c_{j,\sigma} \right) + U_{\text{MFH}} \sum_i \left( n_{i,\uparrow} \langle n_{i,\downarrow} \rangle + n_{i,\downarrow} \langle n_{i,\uparrow} \rangle - \langle n_{i,\uparrow} \rangle \langle n_{i,\downarrow} \rangle \right),$$

where $c_{i,\sigma}^\dagger$ creates a spin σ at lattice site $i$, $n_{i,\sigma}$ is the number operator for spin σ electrons at site $i$, and $\langle n_{i,\sigma}\rangle$ is the average or 'mean-field' spin σ electron density at site $i$ (taking values between 0 and 1). The parameter $t$ is the nearest-neighbour hopping parameter; its corresponding sum is taken over nearest-neighbour pairs. Parameter $U_{\mathrm{MFH}}$ is the Hubbard on-site Coulomb interaction energy parameter, which is related but distinct from the correction $U$ in DFT+$U$. Here, $U_{\mathrm{MFH}}$ represents all electron-electron interactions, in the basis of the kagome lattice, while the $U$ in DFT+$U$ accounts for interactions between Cu $3d$ electrons, not captured accurately by the Hartree and approximate exchange-correlation functionals.

We solved the MFH Hamiltonian numerically with an algorithm written in Python that we summarise here. We used a random electron density $\langle n_{i,\sigma}\rangle$ as an initial ansatz, produced by a symmetric Dirichlet distribution with a high concentration parameter for the spin up and down electron densities separately. Keeping the mean-field electron density constant, the MFH Hamiltonian is divided into spin up and down terms, and solved for its single particle eigenvalues $\epsilon_{j,\sigma}$ and eigenvectors $v_{i,j,\sigma}$ (with $i$ indexing site, $j$ indexing eigenstate and σ indexing spin). We determine the new electron density by filling eigenstates according to the Fermi Dirac distribution:

$$f_{\mathrm{FD}}(E,T) = 1/\left(e^{E/T}+1\right),$$

using a small 'temperature' $T$ (we used $T = 0.1t$). The chemical potential $\mu$ is found by solving for the total electron number $N_e$:

$$N_e = \sum_{j,\sigma} f_{\mathrm{FD}}\left(\epsilon_{j,\sigma}-\mu,T\right).$$

The output electron density is obtained from the eigenvectors by:

$$\langle n_{i,\sigma}^{\mathrm{out}}\rangle = \sum_j f_{\mathrm{FD}}\left(\epsilon_{j,\sigma}-\mu,T\right)\left|v_{i,j,\sigma}\right|^2,$$

and the total energy by:

$$E = \sum_{j,\sigma} f_{FD}(\epsilon_{j,\sigma} - \mu, T)\epsilon_{j,\sigma} - U_{MFH} \sum_{i} \langle n_{i,\uparrow}\rangle\langle n_{i,\downarrow}\rangle.$$

The difference between the input and output electron densities are quantified by the residual:

$$\sqrt{\sum_{i} \left(\langle n_{i,\sigma}^{out}\rangle - \langle n_{i,\sigma}^{in}\rangle\right)^2}.$$

If the residual is sufficiently small, then the calculation has reached self-consistency. Otherwise, we construct a new electron density ansatz by either simple mixing of new and old electron densities or direct inversion of the iterative subspace (DIIS; also known as Pulay mixing[51]), and the process is repeated. Simple mixing sets the new density to be a linear combination of the input and output densities:

$$\langle n_{i,\sigma}\rangle = \alpha\langle n_{i,\sigma}^{out}\rangle + (1-\alpha)\langle n_{i,\sigma}^{in}\rangle,$$

where $\alpha$ is an empirically chosen parameter between 0 and 1. Once simple mixing has reduced the residual < 0.01, we changed to using DIIS[51] using the algorithm of Ref. [52], until the residual was < $10^{-4}$.

For MFH calculations of magnetic moments in the kagome lattice (Fig. 4f), we used a 6×6 supercell with periodic boundary conditions and a 4×4 Monkhorst-Pack $k$-point grid. We took energy units of $t$. To search for the ground state, we sampled different initial electron densities, stepping through ratios of spin up to spin down electrons, and trialling several different random initial configurations for each initial magnetization. Among these trials, the one with the lowest energy after converging to self-consistency was kept and its data reported.

For a direct comparison with the DFT results, we also performed calculations with a primitive unit cell (3 kagome sites) with periodic boundary conditions while sampling the Brillouin zone with a 25×25 Monkhorst-Pack $k$-point grid. We fit the DFT spin configuration on the DCA molecules to the results of the MFH model to find the Hubbard $U_{MFH}$ which minimised the difference between DFT and MFH spin densities (as measured by the root-mean-square difference). The DFT spin density and a uniform charge density was used as the initial configuration for the MFH model in this fitting calculation. Note that the greater configurational freedom for the supercell compared to the primitive unit cell results in an enhanced average local magnetic moment (Fig. 4c) and a richer magnetic configuration (Fig. 4f; SI section S9).

The parameters of interest were the magnetic moments on each kagome site $i$:

$$m^{(i)} = \langle n_{i,\uparrow} \rangle - \langle n_{i,\downarrow} \rangle,$$

the net magnetization:

$$\langle m \rangle = \sum_i \left( \langle n_{i,\uparrow} \rangle - \langle n_{i,\downarrow} \rangle \right) / N_{\text{sites}},$$

where $N_{\text{sites}}$ is the number of sites in the lattice, and the average magnitude of the local magnetic moment on each kagome lattice site:

$$\sqrt{\langle m^2 \rangle} = \sqrt{\left( \sum_i \left( \langle n_{i,\uparrow} \rangle - \langle n_{i,\downarrow} \rangle \right)^2 \right) / N_{\text{sites}}}.$$

**STM Imaging Simulation**

The simulated STM image in Fig. 4d was derived from DFT+$U$ calculations using the Tersoff-Hamann approximation as implemented in HIVE-STM[53], integrating over states from –100 meV to the Fermi level.

# Acknowledgements

D.K., Y.Y. and N.V.M. acknowledge funding support from the Australian Research Council (ARC) Centre of Excellence in Future Low-Energy Electronics Technologies (CE170100039). A.S. acknowledges funding support from the ARC Future Fellowship scheme (FT150100426). B.F. and N.M. gratefully acknowledge the computational support from National Computing Infrastructure and Pawsey Supercomputing Facility. B. F. and B.L. are supported through Australian Government Research Training Program (RTP) Scholarships.

# Supplementary Information

# Manifestation of strongly correlated electrons in a 2D kagome metal-organic framework

Dhaneesh Kumar[†, §], Jack Hellerstedt[†, §], Bernard Field[†, §], Benjamin Lowe[†, §], Yuefeng Yin[§, ±, †], Nikhil V. Medhekar[§, ±] and Agustin Schiffrin[†, §, *]

[†]School of Physics & Astronomy, Monash University, Clayton, Victoria 3800, Australia

[§]ARC Centre of Excellence in Future Low-Energy Electronics Technologies, Monash University, Clayton, Victoria 3800, Australia

[±]Department of Materials Science and Engineering, Monash University, Clayton, Victoria 3800, Australia

* agustin.schiffrin@monash.edu

## Table of Contents

# S1. DCA$_3$Cu$_2$ kagome metal-organic framework on Ag(111) vs Cu(111): structural disorder

The atomically thin DCA$_3$Cu$_2$ kagome structure on Ag(111) is partially disordered, *i.e.*, not perfectly crystalline, as seen in Figs. S1a-d. From atomically and chemical-bond resolved nc-AFM imaging [with tip functionalized with carbon monoxide (CO); see Methods], the DCA$_3$Cu$_2$ kagome metal-organic framework (MOF) on Ag(111) was found to be incommensurate with the underlying substrate, with a 13.0 ± 1.0° offset between MOF and Ag(111) crystalline axes (Figs. S1a-d, Fig.

S2a), and with MOF primitive unit cell vectors $\mathbf{a_1}$ and $\mathbf{a_2}$ such that $\|\mathbf{a_1}\| = \|\mathbf{a_2}\| = 2.045 \pm 0.015$ nm and $\sphericalangle(\mathbf{a_1}, \mathbf{a_2}) = 60 \pm 1°$.

Similarly, we synthesized a DCA$_3$Cu$_2$ kagome MOF on Cu(111) in ultrahigh vacuum (UHV; base pressure 4.5 x 10$^{-10}$ mbar) by depositing thermally evaporated DCA onto clean Cu(111) held at room temperature[1,2]. In these conditions, DCA molecules coordinate with Cu adatoms from the Cu(111) substrate, forming a 2D DCA$_3$Cu$_2$ kagome structure similar to DCA$_3$Cu$_2$ on Ag(111), with quasi-identical primitive unit cell vectors (norm and relative angle). Unlike DCA$_3$Cu$_2$/Ag(111), DCA$_3$Cu$_2$ kagome on Cu(111) is perfectly ordered and commensurate with the underlying Cu(111) substrate (Figs. S1e-h, Fig. S2b).

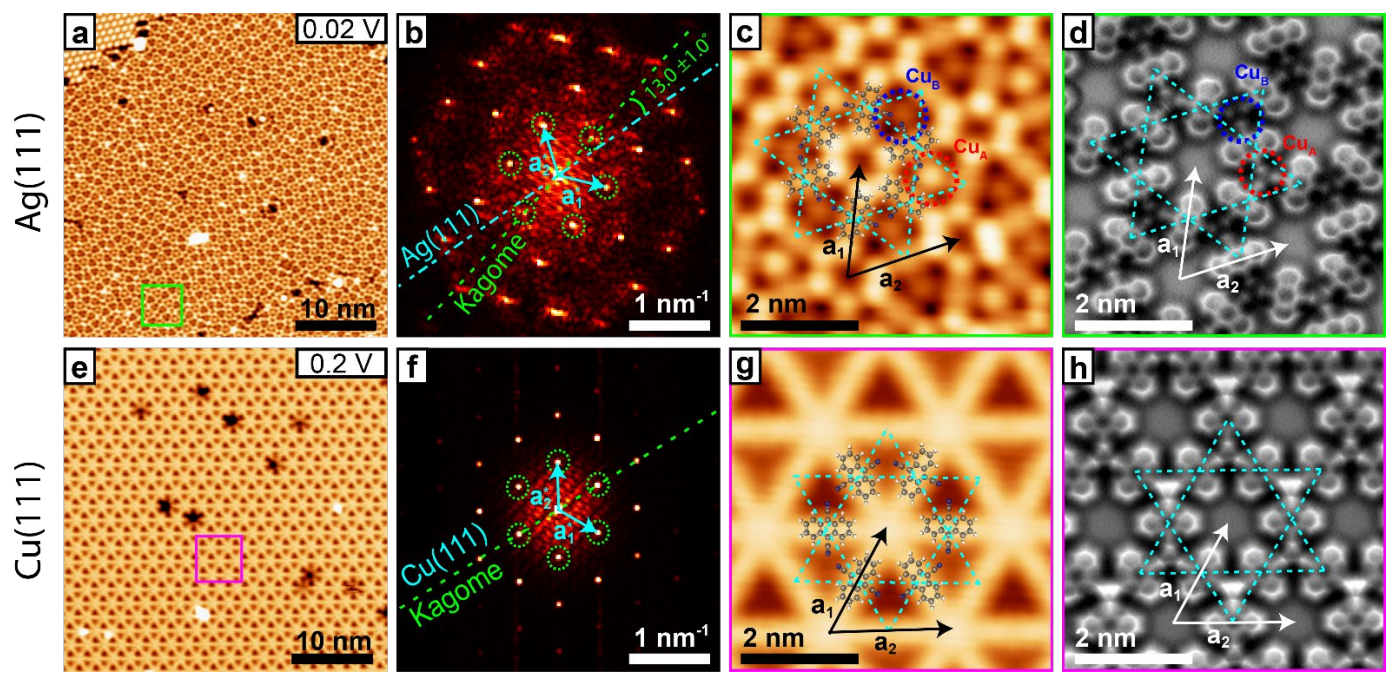

**Figure S1. STM & nc-AFM imaging of 2D DCA$_3$Cu$_2$ kagome MOF on Ag(111) and Cu(111). a,** STM image ($V_b$ = –20 mV, $I_t$ = 50 pA) of DCA$_3$Cu$_2$ kagome MOF on Ag(111). **b,** Fourier transform (FT; |k| = 1/wavelength) of STM image in (a). **c, d,** STM (c; $V_b$ = –20 mV, $I_t$ = 50 pA), and nc-AFM (d; CO-tip; tip retracted 0.3 Å with respect to STM set point $V_b$ = 3 mV, $I_t$ = 150 pA, adjusted on top of DCA molecule centre) images of green-framed region in (a). Reproduced from Fig. 1 of main text. **e-h,** Same as (a)-(d), for DCA$_3$Cu$_2$/Cu(111). Unlike DCA$_3$Cu$_2$/Ag(111), the kagome MOF on Cu(111) is perfectly ordered (see FT's), commensurate and perfectly aligned with the underlying Cu(111) surface:

$\begin{pmatrix} \mathbf{a_1} \\ \mathbf{a_2} \end{pmatrix} = \begin{pmatrix} 8 & 0 \\ 0 & 8 \end{pmatrix} \begin{pmatrix} \mathbf{v_1} \\ \mathbf{v_2} \end{pmatrix}$, where $\{\mathbf{v_1}, \mathbf{v_2}\}$ are the Cu(111) surface primitive unit cell vectors. Set point of $V_b$ = 200 mV, $I_t$ = 50 pA used in (e) and (g); for (h), tip retracted 0.3 Å with respect to STM set point $V_b$ = 2 mV, $I_t$ = 200 pA, adjusted on top of DCA molecule centre.

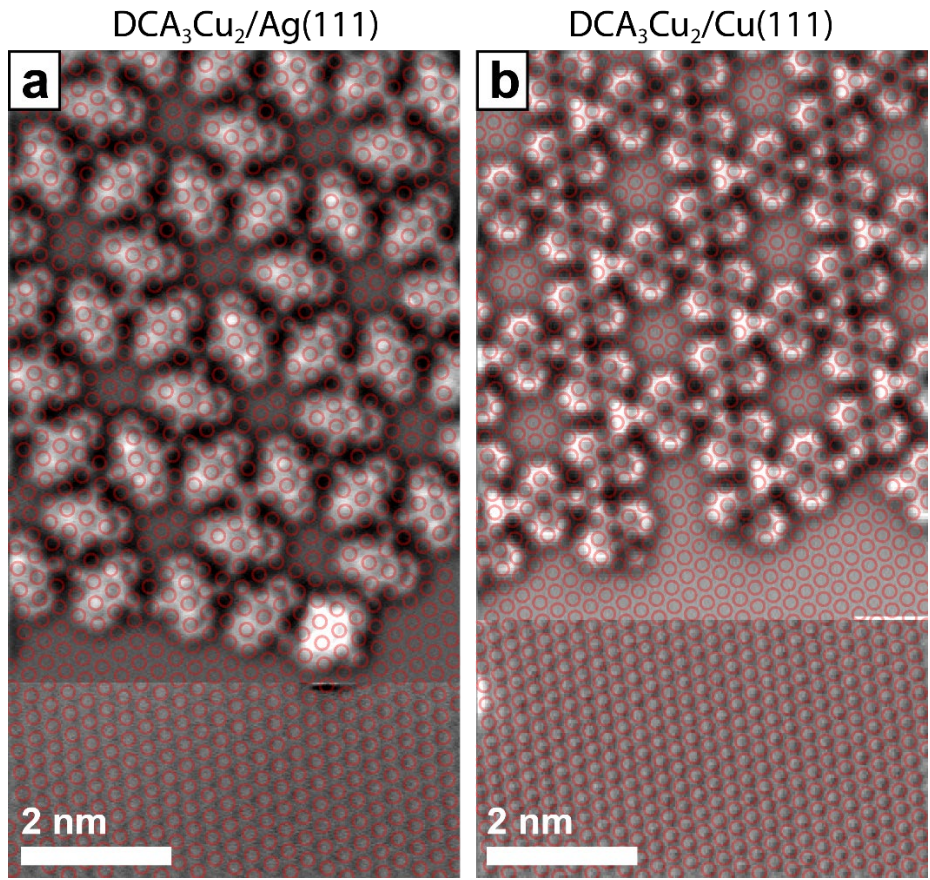

**Figure S2. CO-tip nc-AFM imaging: registration of $DCA_3Cu_2$ on Ag(111) and Cu(111). a,** $DCA_3Cu_2$ on Ag(111). **b,** $DCA_3Cu_2$ on Cu(111). Bottom: atomically resolved bare noble metal surface. Red circles indicate positions of surface atoms. The $DCA_3Cu_2$ structure is incommensurate with the underlying Ag(111) substrate but perfectly commensurate with the Cu(111) substrate. Top: tip retracted 0.3 Å with respect to STM set point $V_b$ = 3 mV, $I_t$ = 0.2 nA, adjusted on top of DCA molecule centre. Bottom: tip height stabilised at STM set point $V_b$ = 3 mV, $I_t$ = 2.5 nA, adjusted on top of DCA molecule centre.

In order to quantitatively compare the degree of disorder between $DCA_3Cu_2$/Ag(111) and $DCA_3Cu_2$/Cu(111), and given that both kagome systems have quasi-identical primitive unit cells and lattice constants, we extracted the Cu atom positions for both kagome systems on Ag(111) and Cu(111) (using MATLAB for image processing; Figs. S3a, b) and then derived the next-nearest-neighbour distances, $r_{NNN}$, for all Cu atom pairs for both systems. Figure S3c shows histograms of $r_{NNN}$ for DCA$_3$Cu$_2$/Ag(111) (blue) and DCA$_3$Cu$_2$/Cu(111) (black). A normal distribution fit for both histograms shows that the average $r_{NNN}$ for both systems are almost identical [$\langle r_{NNN} \rangle$ = 2.039 and 2.043 nm for DCA$_3$Cu$_2$ on Ag(111) and Cu(111), respectively]. However, the standard deviation, $\sigma$, of the distribution of $r_{NNN}$ differs significantly between the two systems ($\sigma$ = 0.088 and 0.016 nm for DCA$_3$Cu$_2$ on Ag(111) and Cu(111) respectively), with the larger width for DCA$_3$Cu$_2$/Ag(111) reflecting the disordered structure compared to the perfectly crystalline DCA$_3$Cu$_2$/Cu(111).

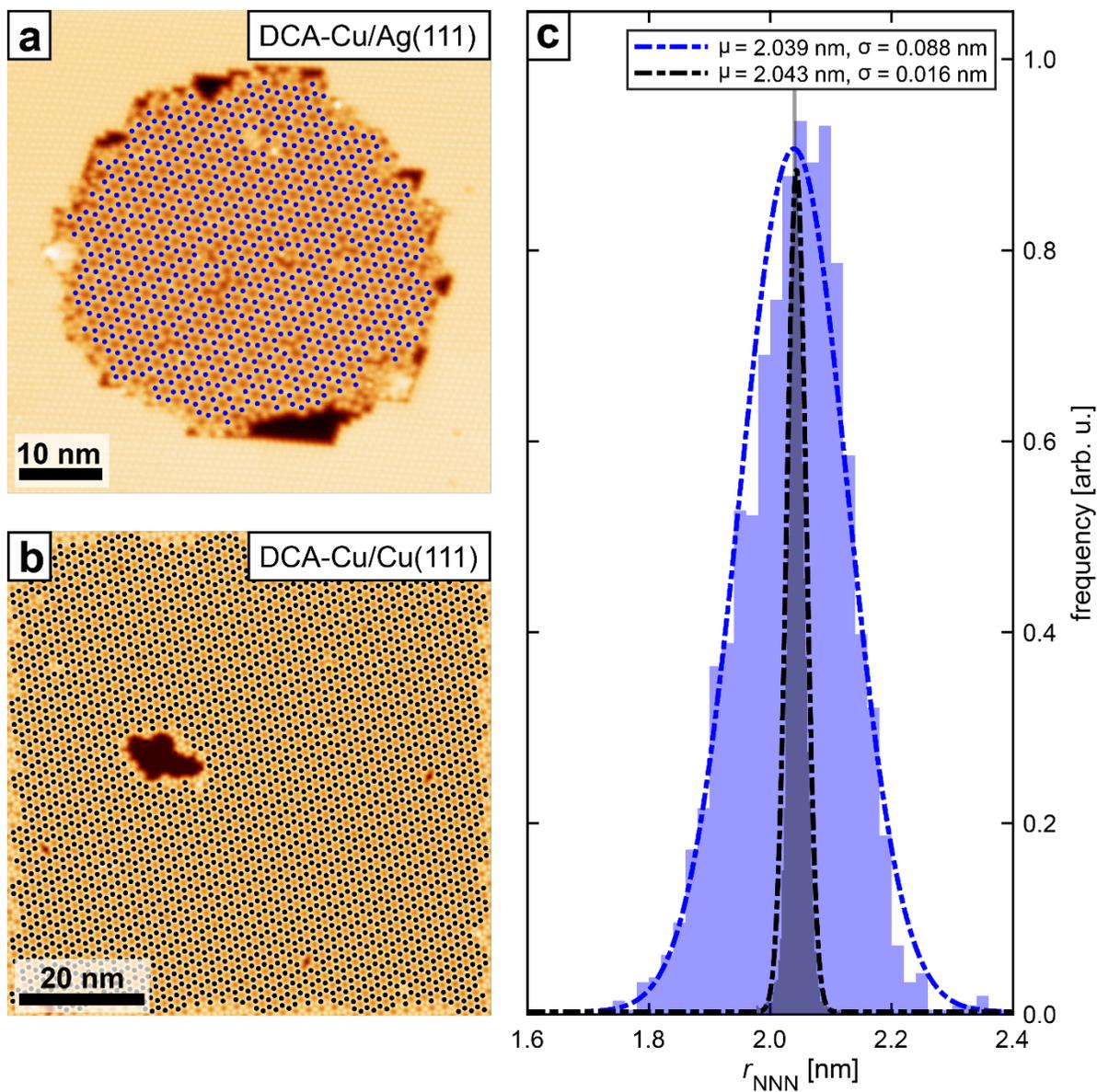

**Figure S3. Cu atom next-nearest-neighbour distances $r_{NNN}$ in DCA$_3$Cu$_2$ on Ag(111) & Cu(111). a-b**, Constant current STM imaging of DCA$_3$Cu$_2$ MOF on Ag(111) and Cu(111), respectively ($V_b$ = 1.0 V, $I_t$ = 50 pA). The Cu atoms' positions (blue & black dots) in both systems were extracted using a *MATLAB* image processing algorithm. **c**, Histogram of the Cu atoms' next-nearest-neighbour distances, $r_{NNN}$, for DCA$_3$Cu$_2$ on Ag(111) (blue) and for DCA$_3$Cu$_2$ on Cu(111) (black). Calculated means μ are similar for both systems; standard deviation σ for DCA$_3$Cu$_2$/Ag(111) system is ~5.5 times larger than for DCA$_3$Cu$_2$/Cu(111).

As mentioned in the main text, we observe two types of Cu atoms within the MOF on Ag(111) (Cu$_A$ and Cu$_B$ in Figs. S1c-d), with different STM and nc-AFM apparent heights. In order to gain insight into the nature of these apparent height differences, we measured the (constant-current) STM apparent height difference, $\Delta z_{app}^{(STM)}$, between Cu$_A$ and Cu$_B$ atoms on Ag(111) (see Figs. S4a-b), for different bias voltages (Fig. S4c). We find a bias-voltage-averaged $\langle \Delta z_{app}^{(STM)} \rangle$ of 21.3 pm (standard deviation: 3.6 pm). We also measured the nc-AFM (with CO-tip) frequency shift, $\Delta f$, as a function of relative tip-sample distance, $z_{rel}$, for both Cu$_A$ and Cu$_B$ atoms (Fig. S4d). The values of $z_{rel}$ associated with the minima of $\Delta f(z_{rel})$ allow us to estimate an nc-AFM apparent height difference, $\Delta z_{app}^{(ncAFM)}$, between Cu$_A$ and Cu$_B$ of ~17 pm, with Cu$_A$ having a higher adsorption height, consistent with $\Delta z_{app}^{(STM)}$. Note that $\Delta z_{app}^{(ncAFM)}$ varies slightly depending on which Cu$_A$ and Cu$_B$ atoms are considered throughout the MOF. From this we conclude that the observed difference in STM and nc-AFM appearance of Cu atoms in DCA$_3$Cu$_2$/Ag(111) is the result of a difference in real adsorption height of the Cu atoms, of ~19 pm, due to the varying adsorption sites of the Cu atoms on Ag(111) [given by the incommensurability of the DCA$_3$Cu$_2$ structure with the Ag(111) substrate].

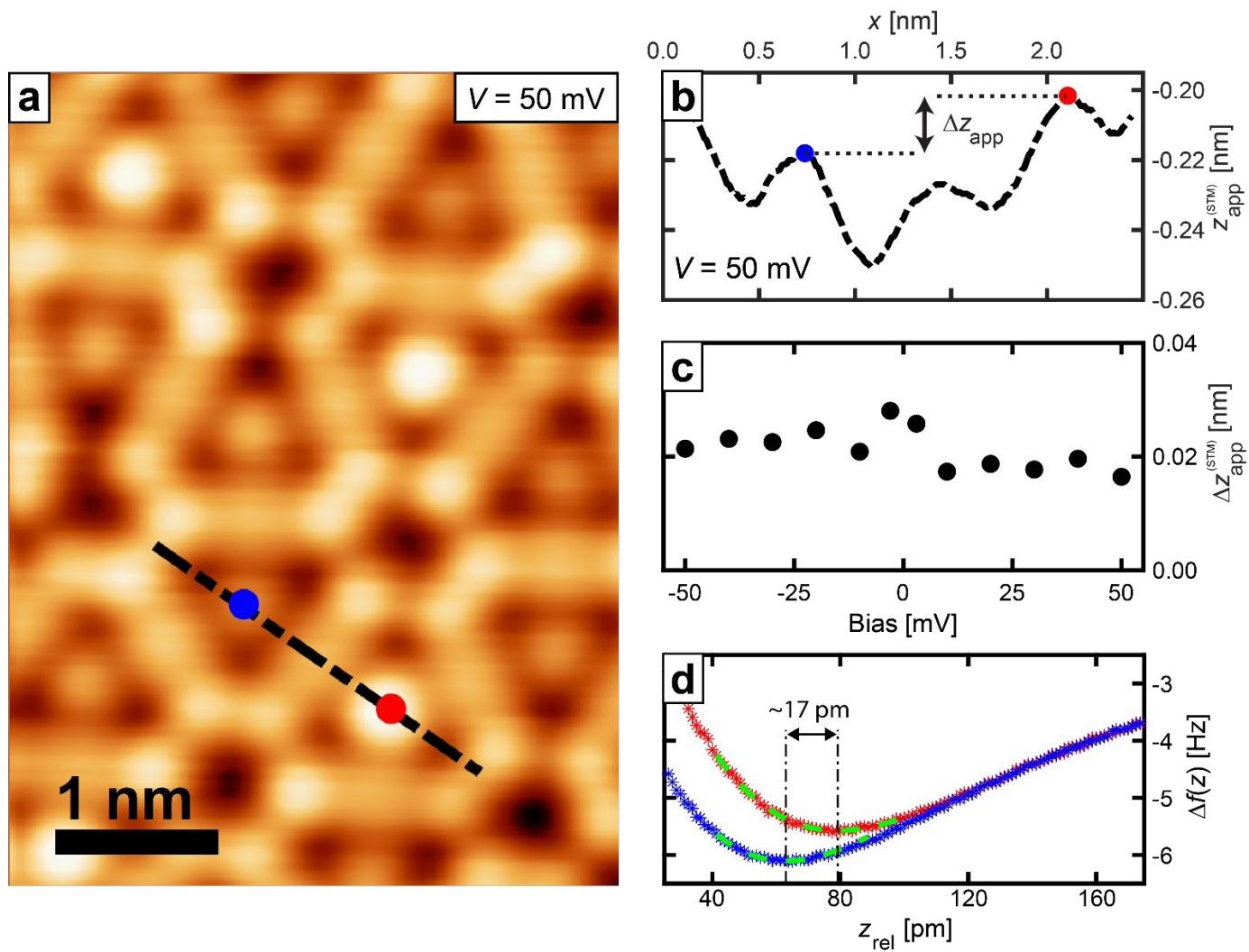

**Figure S4. Adsorption height difference between Cu$_A$ and Cu$_B$ atoms in DCA$_3$Cu$_2$/Ag(111). a**, Constant-current STM imaging of DCA$_3$Cu$_2$/Ag(111) ($V_b$ = 50 mV, $I_t$ = 50 pA). Red circle: Cu$_A$. Blue: Cu$_B$. **b**, Constant-current STM apparent height, $z_{app}^{(STM)}$, along black dashed line in (a): $\Delta z_{app}^{(STM)} = z_{app}^{(STM, Cu_A)} - z_{app}^{(STM, Cu_B)}$. **c**, Difference in STM apparent height, $z_{app}^{(STM)}$, between Cu$_A$ and Cu$_B$, as a function of bias voltage. **d**, Nc-AFM frequency shift as a function of relative tip-sample distance, $\Delta f(z_{rel})$, for Cu$_A$ (red curve) and Cu$_B$ (blue curve). Green dashed curves: 3$^{rd}$ order polynomial fit. The difference in $z_{rel}$ associated with the $\Delta f(z_{rel})$ minima is a measure of the difference in adsorption height between Cu$_A$ and Cu$_B$: $z_{app}^{(ncAFM)} \approx 17$ pm ($z_{rel} = 0$ pm corresponds to an STM set point of $V_b$ = 3 mV, $I_t$ = 150 pA, adjusted on top of DCA molecule centre).

## S2. d*I*/d*V* STS on DCA$_3$Cu$_2$/Ag(111): confinement of Shockley surface state

In addition to near-Fermi d*I*/d*V* features (Fig. 2 of the main text), d*I*/d*V* spectra taken for bias voltage $V_b$ > 0.1 V at high-symmetry points of the DCA$_3$Cu$_2$ kagome structure on Ag(111) show additional peak-like features (Fig. S5a). Based on previous work[1,3] on DCA$_3$Cu$_2$/Cu(111), we investigate here whether these features are due to the (partial) confinement of the Ag(111) Shockley surface state[4] by the kagome structure, by simulating the local density of states (LDOS) of the DCA$_3$Cu$_2$/Ag(111) system according to the Electron Plane Wave Expansion (EPWE) model[5]. This model assumes a 2D piecewise constant periodic potential resulting from the DCA$_3$Cu$_2$ kagome structure, which results in scattering and confinement of the Ag(111) Shockley surface state. Figure S5d shows a region of this potential (modelling the DCA$_3$Cu$_2$/Ag(111) structure shown in Fig. S5c) where the height of the potential barrier given by Cu atoms is $V_{Cu}$ = 1.5 V (red regions), that given by DCA molecule centres is $V_{mol}$ = 0.15 V (cyan), that given by DCA molecule anthracene extremities is $V_{end}$ = 0.15 V (green), and the background potential given by Ag(111) is $V_{Ag}$ = 0 V (blue). The primitive unit cell of this 2D potential is given by the DCA$_3$Cu$_2$ MOF unit cell measured by STM ($\|\mathbf{a_1}\| = \|\mathbf{a_2}\| = 2.045$ nm with $\sphericalangle(\mathbf{a_1}, \mathbf{a_2}) = 60°$; see Fig. S1). Figure S5b shows the simulated LDOS at high-symmetry points of the DCA$_3$Cu$_2$ kagome structure (binning of 30 mV). For 0.1 V < $V_b$ < 1.2 V the features in our experimental d*I*/d*V* spectra are reproduced qualitatively by our simulated LDOS (Figs. S5a, b). Figures S5e, g, i show experimental d*I*/dV maps at $V_b$ = 0.35, 0.58 and 1.2 V, obtained by acquiring and averaging multiple *I*(*V*) curves followed by numerical derivation of the averaged *I*(*V*) curves (with respect to bias voltage) at each tip position. Simulated LDOS maps (Figs. 5f, h, j) reproduce qualitatively our experimental d*I*/d*V* maps, in particular the high (low) LDOS in the MOF pore (at kagome sites, respectively) for 0.1 < $V_b$ < 0.5 V, and the high (low) LDOS at the kagome sites (MOF pore, respectively) for at $V_b$ > 0.5 V. Note that the EPWE model energy scale offset (determined by the Shockley surface state energy onset) was fixed such that the simulated LDOS in Fig. S5b matches the d$I$/d$V$ spectra in Fig. S5a. Differences between experimental and simulated data, in particular for $V_b$ > 0.5 V, could be explained by the structural periodicity of the kagome structure assumed in the model (in contrast with the experimentally observed partial disorder of DCA$_3$Cu$_2$/Ag(111); Figs. S1 and S2; Fig. 1a of main text), as well as by the assumed piecewise constant potential (in contrast with the real continuous potential).

From this qualitative agreement between our experimental and simulated data, we conclude that the d$I$/d$V$ features for $V_b$ > 0.1 V are mostly due to confinement of the Ag(111) Shockley surface state by the DCA$_3$Cu$_2$ kagome MOF, and are not related to any intrinsic DCA$_3$Cu$_2$ kagome electronic state.

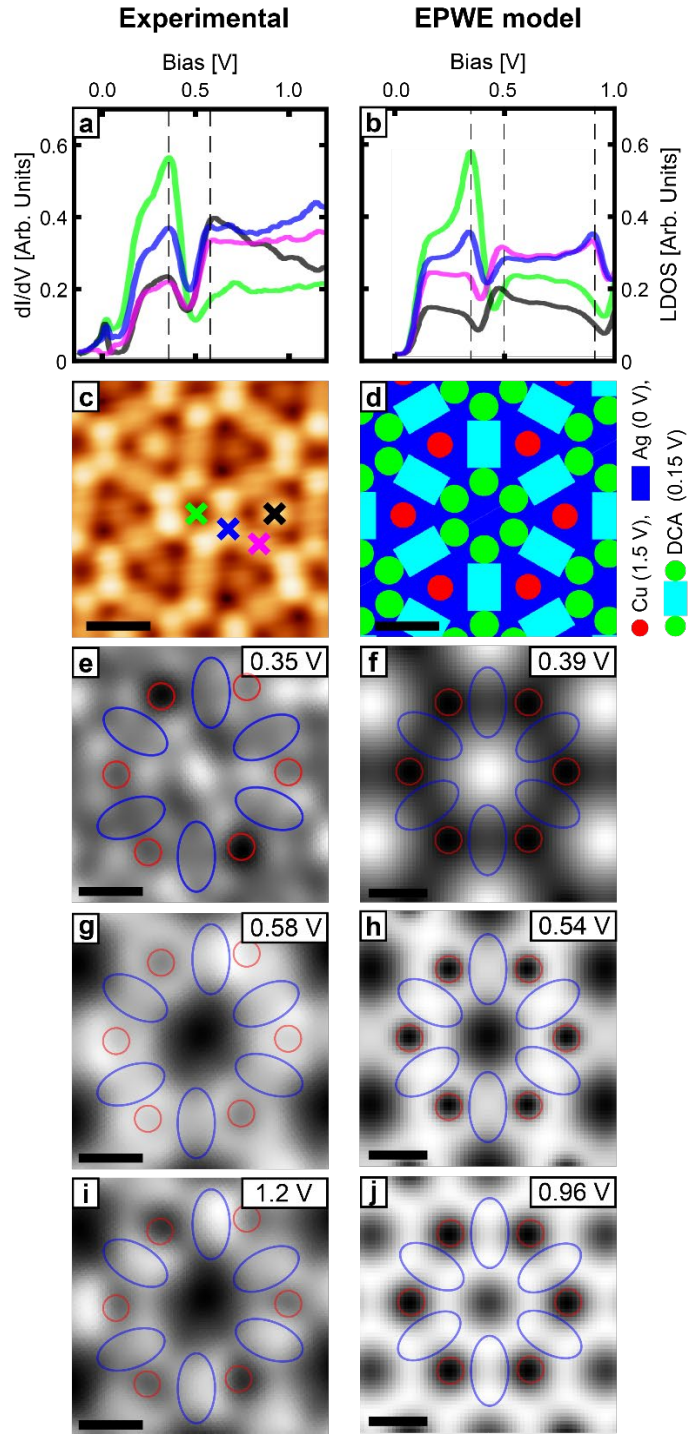

**Figure S5. d*I*/d*V* maps and simulated LDOS of DCA$_3$Cu$_2$/Ag(111) *via* Electron Plane Wave Expansion (EPWE) model. a**, Differential conductance (d*I*/d*V*) STS spectra at high-symmetry points [shown by crosses in (c)] showing features at $V_b$ = 0.35, 0.58 and 1.2 V (set point: $V_b$ = –0.3 V, $I_t$ = 20 pA). **b**, LDOS at high-symmetry points simulated by EPWE model [2D periodic piecewise constant potential shown in (d)] showing features similar to experimental data

S11

in (a). Energy scale was offset to best match with experimental data. **c**, Constant-current STM image of the DCA$_3$Cu$_2$/Ag(111) kagome structure (set point: $V_b$ = –20 mV, $I_t$ = 50 pA). **d**, 2D periodic piecewise constant potential used to model the DCA$_3$Cu$_2$ kagome structure in (c). **e**, **g**, **i**, Experimental d$I$/d$V$ maps (set point $V_b$ = –0.3 V, $I_t$ = 20 pA), at indicated bias voltages. Blue ellipses and red circles indicate DCA molecules and Cu atoms, respectively. **f**, **h**, **j**, LDOS simulated by EPWE model at indicated energies, showing good agreement with experiment, and confirming that most d$I$/d$V$ features at bias voltages >~0.1 V are due to scattering and confinement of Ag(111) Shockley surface state. Scale bars: 1 nm. Experimental dI/dV spectra in (a) and maps in (e), (g), (i) obtained via numerical derivation (with respect to bias voltage) of averaged $I(V)$ curves acquired at each tip position.

## S3. d$I$/d$V$ STS along high-symmetry lines of DCA$_3$Cu$_2$/Ag(111)

We acquired differential conductance (d$I$/d$V$) spectra along two high symmetry lines of the DCA$_3$Cu$_2$ kagome system on Ag(111): (i) along the DCA long anthracene axis (Fig. S6b); and (ii) along a Cu$_A$-Cu$_B$ axis (Fig. S6c). We note that for bias voltages |$V_b$| > 43 mV, the d$I$/d$V$ signal is larger at the DCA molecular centre than at the DCA anthracene extremities or Cu atoms. These features are related to molecular vibrational modes and vibrationally assisted Kondo effect. For |$V_b$| < 43 mV, d$I$/d$V$ signal is greater at anthracene extremities and Cu atoms than at the DCA molecular centres; these features are related to the zero-bias LDOS Kondo resonance.

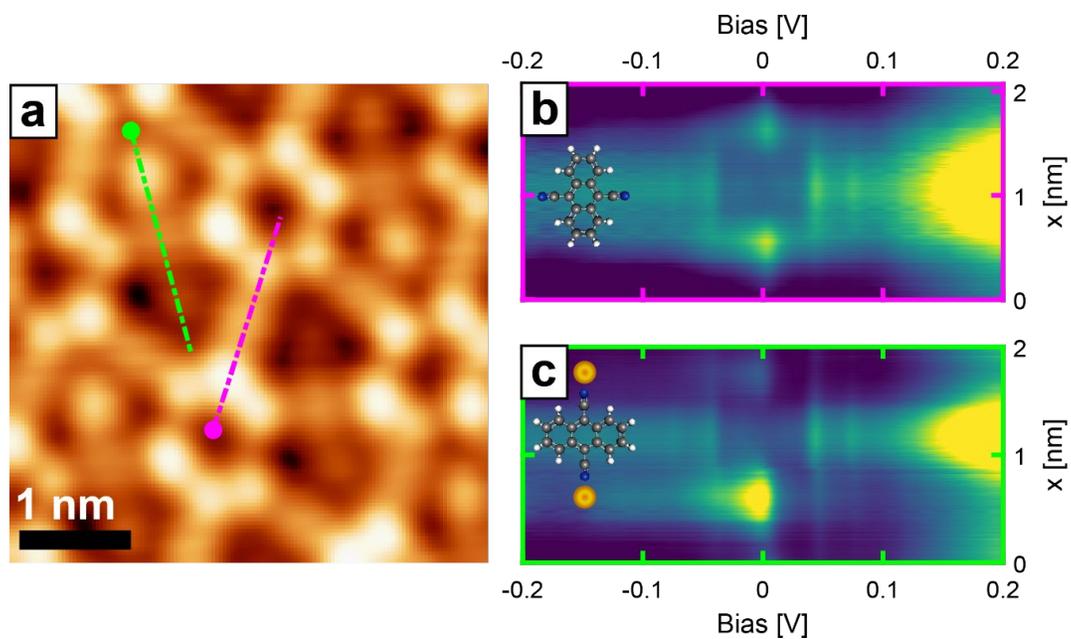

**Figure S6. Differential conductance (d*I*/d*V*) measurements along high-symmetry lines of the DCA$_3$Cu$_2$ kagome structure on Ag(111). a**, Constant-current STM image of DCA$_3$Cu$_2$ kagome structure on Ag(111) ($V_b$ = 20 mV, $I_t$ = 50 pA). **b**, d*I*/d*V* spectra along DCA anthracene axis [magenta dashed line in (a) with disk marker indicating $x$ = 0 nm; set point: $V_b$ = –200 mV, $I_t$ = 0.4 nA]. **c**, d*I*/d*V* spectra along Cu$_A$-Cu$_B$ axis [green dashed line in (a); set point: $V_b$ = –200 mV, $I_t$ = 0.4 nA]. For |$V_b$| > 43 mV, d*I*/d*V* signal is larger at DCA centre (vibrational modes and vibrationally assisted Kondo effect) than at Cu atoms or anthracene extremities. For |$V_b$| < 43 mV, d*I*/d*V* is larger at Cu atoms and at anthracene extremities (zero-bias Kondo effect). These d*I*/d*V* spectra were obtained by acquiring and averaging multiple *I*(*V*) curves at each tip position, followed by numerical derivation of the average *I*(*V*) curves (with respect to bias voltage).

## S4. Near-Fermi d*I*/d*V* maps

Figure S7 shows d*I*/d*V* maps of DCA$_3$Cu$_2$ on Ag(111) for bias voltages –100 < $V_b$ < 100 mV. The maps for |$V_b$| = 100 (Figs. S7b, f) and 73 mV (Figs. S7c, g) display higher intensity at molecule centres compared to anthracene extremities and Cu atom sites, related to molecular vibrational modes (see section S6 below). The maps at $V_b$ = –100 and –73 mV also show significant intensity at the pore centres, related to confinement of the Ag(111) Shockley surface state (section S2

above). The maps for –43 < $V_b$ < 43 mV (Figs. S7d, e, h-j) reveal higher intensity at the anthracene extremities and Cu atoms than at the DCA centres, related to the zero-bias Kondo peak.

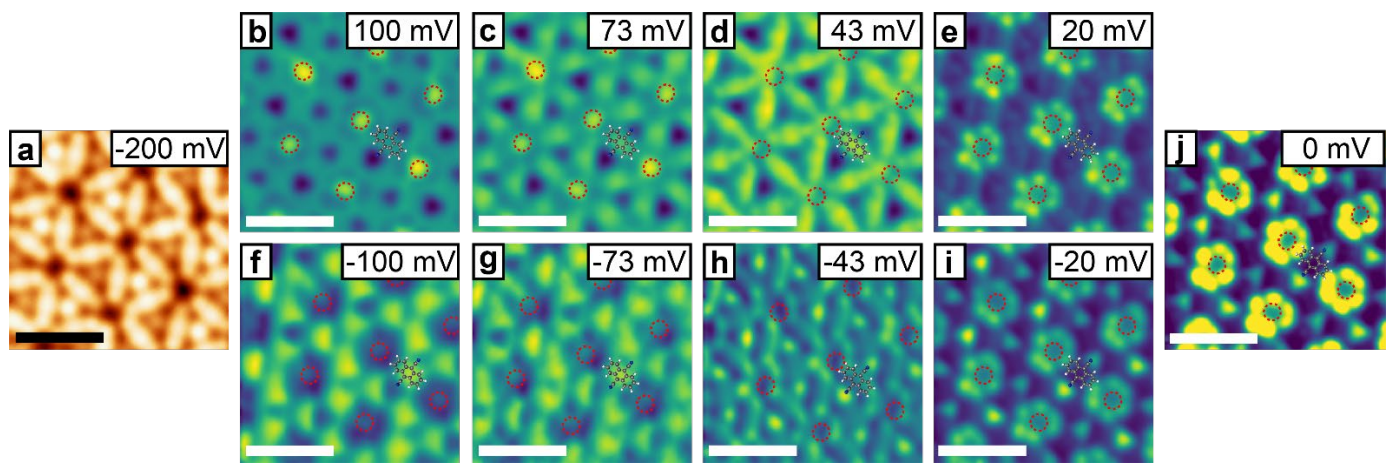

**Figure S7.** d$I$/d$V$ maps of DCA$_3$Cu$_2$/Ag(111) for bias voltages -100 < $V_b$ < 100 mV. **a**, Constant-current STM image ($V_b$ = –200 mV, $I_t$ = 500 pA). **b-j**, d$I$/d$V$ maps of region in (a), for $V_b$ indicated in each panel. See Methods section in main text for measurement details. Scale bars: 2 nm. DCA structure superimposed. Red dashed circles indicate kagome MOF pores.

## S5. Tip-sample distance-dependent d$I$/d$V$ STS

To rule out any possible tip-related effects on the electronic structure of our system (e.g., tip-induced changes of molecular confirmation[6], electronic state energy level shifting due to tip-induced gating[7]), and in particular in the ZBP, we performed tip-sample distance-dependent d$I$/d$V$ measurements. Figures S8a-c show d$I$/d$V$ spectra acquired on Cu$_A$, Cu$_B$, and DCA anthracene extremity sites at different tip heights relative to an STM set point ($V_b$ = –300 mV, $I_t$ = 1.5 nA). The d$I$/d$V$ signal intensity decreases monotonically with increasing tip-sample distance, with the shape of the near-Fermi d$I$/d$V$ features remaining qualitatively unaltered. Figure S8d shows the ZBP peak intensity (i.e., d$I$/d$V$ maximum normalized by the d$I$/d$V$ signal at $V_b$ = –300 mV) for Cu$_A$, Cu$_B$, and DCA anthracene extremity sites as a function of tip-sample distance, revealing only a marginal decrease in peak intensity as the tip-sample distance in increased. We can explain this very small decrease in peak intensity by a decrease in the background d$I$/d$V$ signal. From this, we conclude that the electronic properties of DCA$_3$Cu$_2$/Ag(111) are not affected by the tip.

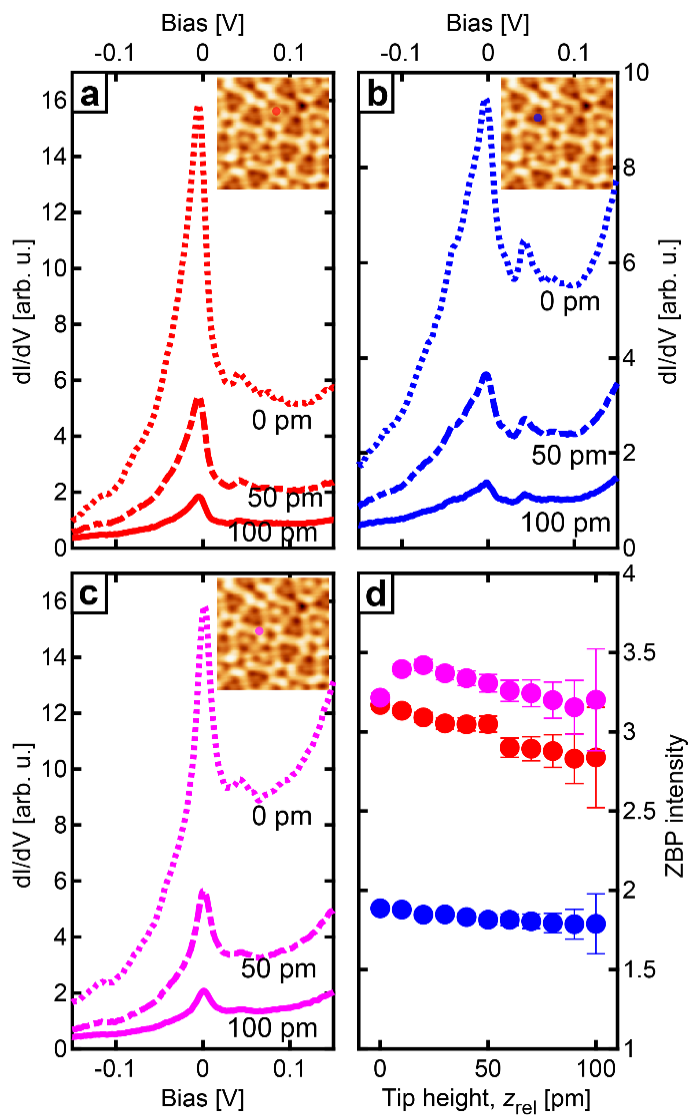

**Figure S8. Tip-sample distance-dependent d$I$/d$V$ spectra at high symmetry points for DCA$_3$Cu$_2$/Ag(111). a-c**, d$I$/d$V$ spectra at a Cu$_A$ (a), Cu$_B$ (b), and DCA anthracene extremity sites (c) for various increasing tip-sample distances relative to an STM set point of $V_b$ = –300 mV, $I_t$ = 1.5 nA. **d**, d$I$/d$V$ ZPB intensity (d$I$/d$V$ peak maximum normalized by the d$I$/d$V$ signal at $V_b$ = –300 mV) as a function of tip-sample distance, for Cu$_A$ (red), Cu$_B$ (blue) and DCA anthracene extremity (magenta) sites. Insets: constant current STM images (set point: $V_b$ = 20 mV, $I_t$ = 100 pA).

## S6. Near-Fermi dI/dV STS

Details of fitting procedure

We fit all dI/dV spectra (at $Cu_A$, $Cu_B$, DCA anthracene extremity and DCA molecular centre sites) with a non-linear least squares algorithm (*MATLAB*), with the fitting function:

$$(dI/dV)_\text{fit}(V,T) = \int_{-\infty}^{\infty} \rho_\text{fit}(E,T) \frac{d}{dV} f_\text{FD}(E-eV,T)\, dE$$

where $T$ is the temperature of the system, $V$ is the bias voltage, $\rho_\text{fit}(E,T)$ accounts for the local density of states (LDOS) at a particular site and $f_\text{FD}(E,T)$ is the Fermi-Dirac distribution. The derivative of $f_\text{FD}(E,T)$ with respect of $V$, $\frac{d}{dV} f_\text{FD}(E-eV,T)$, accounts for trivial thermal broadening of the electronic energy distribution[8].

For the $Cu_A$ and $Cu_B$ sites we considered the following $\rho_\text{fit}(V,T)$:

$$\rho_\text{fit}^{(Cu)}(V,T) = \text{Fano}(V,E_0,\Gamma_0,q,A) + \sum_i L(V,E_i,\Gamma_i,A_i) + \text{background}(V)$$

where $\text{Fano}(V,E_0,\Gamma_0,q,A)$ is a Fano line shape[8,9] with fitting parameters $E_0, \Gamma_0, q, A$:

$$\text{Fano}(V,E_0,\Gamma_0,q,A) = \frac{A}{1+q^2} \frac{(\epsilon+q)^2}{\epsilon^2+1}, \quad \epsilon = \frac{eV-E_0}{\Gamma_0},$$

$L(V,E_i,\Gamma_i,A_i)$ are Lorentzians with fitting parameters $E_i, \Gamma_i, A_i$, centered at energies $E_i$:

$$L(V,E_i,\Gamma_i,A_i) = A_i \frac{\frac{1}{4}\Gamma_i}{(V-E_i)^2 + \left(\frac{1}{4}\Gamma_i\right)^2}.$$

and the third term is a bias-dependent background (discussed below). The Lorentzian functions $L(V,E_i,\Gamma_i,A_i)$ fit the dI/dV spectra satellite peaks at ~±43 mV and ±73 mV.

For the DCA anthracene extremities we considered:

$$\rho_\text{fit}^{(\text{lobe})}(V,T) = \sum_i L(V,E_i,\Gamma_i,A_i) + \text{background}(V).$$

Here, the ZBP at the DCA anthracene extremity was fit with a Lorentzian function given its symmetry; note that a Lorentzian is identical to a Fano line shape with $q \to \infty$.

For the DCA molecular centres, we considered:

$$\rho_{\text{fit}}^{(\text{centre})}(V,T) = S(V, E_0, \Theta, A_0, A'_0) + \sum_i L(V, E_i, \Gamma_i, A_i) + \text{background}(V)$$

where:

$$S(V, E_0, \Theta, A_0, A'_0) = A_0 \, \text{erf}\left(\left(\frac{V - E_0}{\Theta}\right) + 1\right) + A'_0 \, \text{erf}\left(1 - \left(\frac{V + E_0}{\Theta}\right)\right).$$

The term $S(V, E_0, \Theta, A_0, A'_0)$ accounts for the step-like features, symmetric in energy with respect to $V_b = 0$, at ~±43. Here $E_0, \Theta, A_0, A'_0$ are free fitting parameters corresponding to the steps' location in energy, step broadening term, and steps' amplitude. The Lorentzian functions $L(V, E_i, \Gamma_i, A_i)$ fit the off-Fermi d*I*/d*V* satellite peaks.

For all $\rho_{\text{fit}}(V,T)$, the following background was considered:

$$\text{background}(V) = A \exp\left(\frac{V - E'}{\sigma}\right) + \text{constant}$$

where $A, E'$ and $\sigma$ are free fitting parameters, accounting to the rise in d*I*/d*V* signal for $V_b > 0.1$ V due to features between 0.2 and 0.4 V related to the confinement of the Ag(111) Shockley surface state by the MOF(see Fig. 2 of main text; section S2 above).

## Temperature dependence of ZBP

From the above fitting procedure, we can obtain the width $\Gamma$ of the ZBP, as a function of temperature, for each site (Fig. S9a). We determined the Kondo temperature, $T_K$, at each site by fitting $\Gamma(T)$ with Eq. (3) of the main text (see Table S1). Note that for $Cu_B$ sites $T_K = 162 \pm 16$ K, similar to that for $Cu_A$ and DCA anthracene extremity sites.

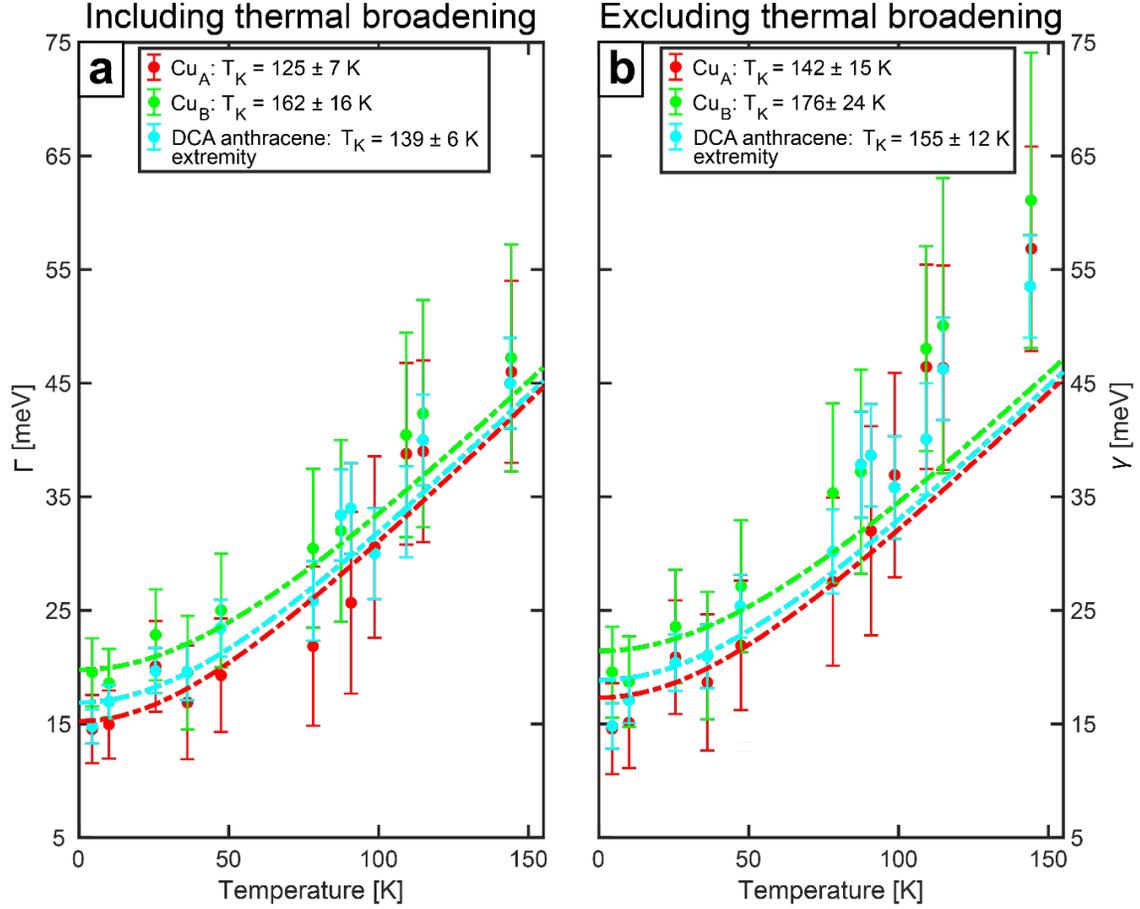

**Figure S9. Broadening of d$I$/d$V$ ZBP as a function of temperature $T$, with and without thermal broadening of the Fermi-Dirac distribution. a, b**, ZBP widths, $\Gamma$ and $\gamma$, as a function of $T$ for Cu$_A$ (red), Cu$_B$ (green) and DCA anthracene extremity (cyan) sites, extracted from Fano (Cu) or Lorentzian (anthracene extremity) line shape fits with (a) and without (b) accounting for Fermi-Dirac distribution thermal broadening. Dashed curves: fits given by Eq. (3) of main text. Data for Cu$_A$ and DCA anthracene extremity sites reproduced from Fig. 3d of main text.

| Location | Shape parameter, $q$ (at 4.4 K) | ZBP energy position [meV] (at 4.4 K) | Kondo temperature, $T_K$ |
|---|---|---|---|
| Cu$_A$ | −3 | 0.9 | 125 ± 7 K |
| Cu$_B$ | −2 | 2.9 | 162 ± 16 K |
| DCA anthracene extremity | >1000000 | 1.2 | 139 ± 6 K |

**Table S1. ZBP fitting parameters.** Parameters for the ZBP obtained from fitting the temperature-dependent d$I$/d$V$ spectra at Cu$_A$, Cu$_B$ and DCA anthracene extremity sites.

S18

Further, note that the absolute value of the Fano parameter, |q|, is greater for $Cu_A$ than for $Cu_B$. Parameter |q| represents the ratio between the rate of direct tunnelling into the Kondo impurity, and the rate of tunnelling into continuum states of the underlying metal substrate. The lower the value of |q| is, the greater the coupling between the Kondo impurity and the substrate[9]. This hints towards the $Cu_A$ atoms being less coupled to the substrate in comparison to the $Cu_B$ atoms. This is consistent with the observed adsorption heights of $Cu_A$ and $Cu_B$, with $Cu_B$ adsorbed closer to Ag(111) (see Section S1). Similar observations have been reported for NTCDA molecules on Ag(111)[10].

To test our d$I$/d$V$ spectra fitting procedure, we have also tried fitting these d$I$/d$V$ spectra without considering thermal broadening of the Fermi-Dirac distribution, that is, by considering a fitting function $(dI/dV)_{\text{fit}}(V,T) = \rho_{\text{fit}}(E,T)$ accounting for the LDOS at a particular site and consisting of the same functions as above. Figure S9b shows the width, $\gamma$, of the ZBP fitting Fano line shape (for Cu atom sites) and of the ZBP fitting Lorentzian (for the DCA anthracene extremity), as a function of temperature $T$, resulting from this fit without thermal broadening of the Fermi-Dirac distribution. As for $\Gamma(T)$ (see above), we fit $\gamma(T)$ with Eq. (3) of the main text (Fig. S9b). While there is agreement between $\gamma(T)$ and this fit at low $T$ (where thermal broadening is negligible), there is a significant deviation at higher $T$ (where thermal broadening is more substantial). Note that the only free fitting parameter in Eq. (3) of the main text is the Kondo temperature, $T_K$, which is constrained by the low-temperature behaviour of $\Gamma(T)$ and $\gamma(T)$ since $\sqrt{2(k_B T_K)^2 + (\pi k_B T)^2} \to \sqrt{2} k_B T_K$ for $T \to 0$, and the slope of the fitting function in Eq. (3) for $T \gg T_K$, $d\left(\sqrt{2(k_B T_K)^2 + (\pi k_B T)^2}\right)/dT \to \pi k_B$ (independent of $T_K$). That is, although the fit for $\gamma(T)$ is not as good as for $\Gamma(T)$, the extracted values of $T_K$ are similar (Fig. S9).

## Spatial dependence of d$I$/d$V$ ZBP

Figure S10 shows d$I$/d$V$ spectra acquired at five different DCA anthracene extremity sites (Fig. S10a) and five different Cu$_A$ sites (Fig. S10b), with the same tip and same acquisition parameters, at 4.4 K. Figure S10c shows the width $\Gamma$ and amplitude of the fit Fano (for Cu$_A$) and Lorentzian (DCA anthracene extremity) line shapes (see fitting procedure above, including thermal broadening of Fermi-Dirac distribution) for the d$I$/d$V$ ZBP at these different sites. The amplitude of the ZBP varies to some degree from one site to another, possibly due to small variations (e.g., thermal drift) away from the optimal tip position. This hindered a reliable quantitative measurement of the temperature dependence of the ZBP amplitude. Note that the latter can be used to independently determine the Kondo temperature and to corroborate the value obtained via the temperature-dependent ZBP broadening, as well as to infer the spin state of the system (*e.g.*, S = 1/2 or S = 1)[11]. On the other hand, $\Gamma$ remains fairly constant for different Cu$_A$ and DCA anthracene extremity sites. This could be explained by the nonlinear relationship between amplitude and width $\Gamma$ of the Fano/Lorentzian line shape (that is, a small variation of the amplitude of these line shapes results in an even smaller relative variation of $\Gamma$), resulting in $\Gamma$ being more robust and less sensitive to the exact probe location (and other factors such as STM set point and atomic-scale tip morphology) in comparison with the line shape amplitude. This justifies our conclusion, based on the temperature dependence of $\Gamma$ and its trend given by Eq. (3) of the main text (Fig. 3d of main text), that the ZBP is unambiguously related to the Kondo effect.

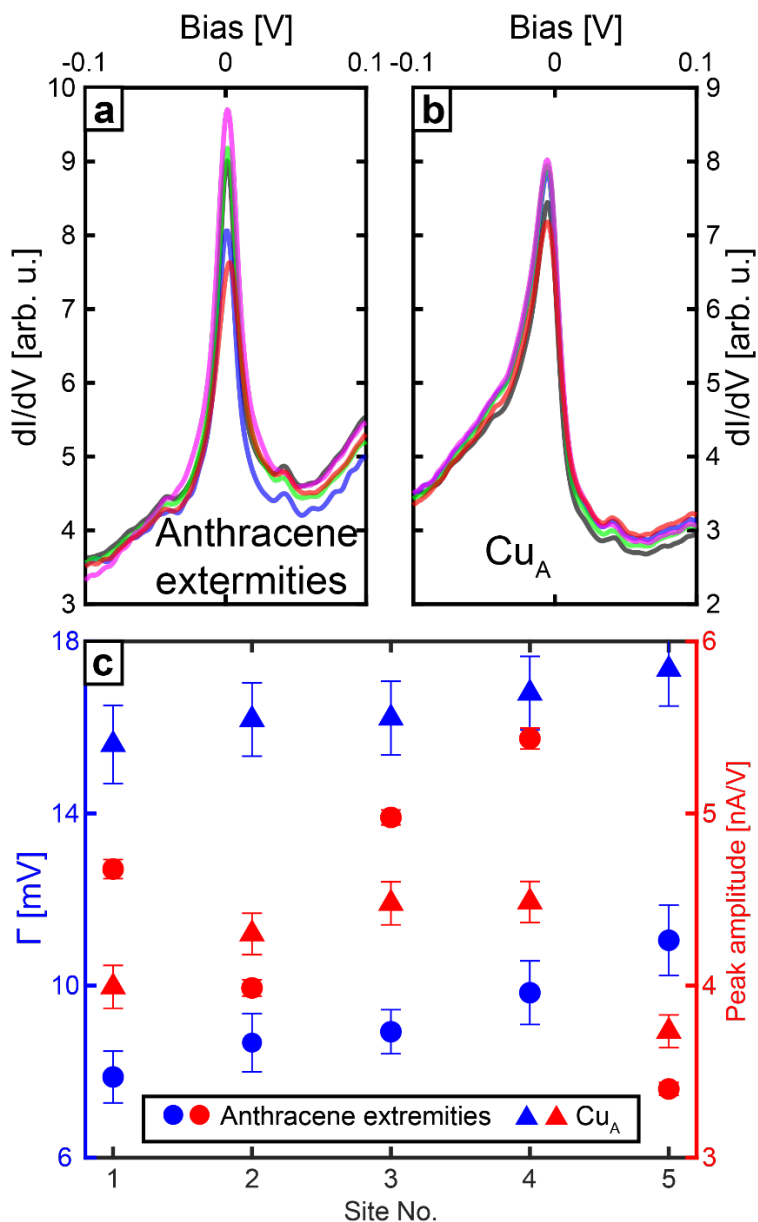

**Figure S10. d*I*/d*V* ZBP spatial dependence. a-b**, d*I*/d*V* spectra at (a) different DCA anthracene extremity and (b) different Cu$_A$ sites (STM set point: $V_b$ = –250 mV, $I_t$ = 1 nA). **c**, Width Γ (blue) and amplitude (red) of fitting Lorentzian and Fano line shapes for ZBP of curves in (a) and (b).

## Off-Fermi d*I*/d*V* satellite peaks and step features

Our fitting procedure above (including thermal broadening of Fermi-Dirac distribution) allowed for the extraction of the energy positions and widths of the fitting Lorentzians for the d*I*/d*V* satellite peaks, as well as the energy positions of the d*I*/d*V* steps at the DCA molecular centre, at 4.4 K (see Table S2). We acquired d*I*/d*V* data at the DCA molecular centre using different tips to ensure that all d*I*/d*V* features are intrinsic to the kagome system and are not tip related. Figure S11 shows d*I*/d*V* data acquired at the DCA molecular centre with four different tips. In addition to the off-Fermi, energy-symmetric satellite peaks observed at ±17.3 mV, ±43.5 mV, and ±75.6 mV, reported in the main text (black dashed lines), we note other subtler features at ±56.3 mV and ±85.8 mV (red dashed lines). Because these features are smaller than the ones at ±17.3 mV, ±43.5 mV, and ±75.6 mV, in the main text we focused on the latter.

| d*I*/d*V* satellite peaks | | d*I*/d*V* step feature | |
|---|---|---|---|
| Peak energy position [meV] (at 4.4 K) | Peak width, $\Gamma_i$ [meV] (at 4.4 K) | Step energy position [meV] (at 4.4 K) | Step broadening term, $\Theta$ [meV] (at 4.4 K) |
| ±17.3 | 8.8 | ±43.5 | 15.0 |
| ±43.5 | 7.3 | | |
| ±56.3 | 9.1 | | |
| ±75.6 | 6.3 | | |
| ±85.8 | 8.6 | | |

**Table S2. Near-Fermi energy-symmetric d*I*/d*V* features fitting parameters.** Parameters for the d*I*/d*V* satellite peaks and step features obtained from fitting the d*I*/d*V* spectra at DCA centre sites at 4.4 K.

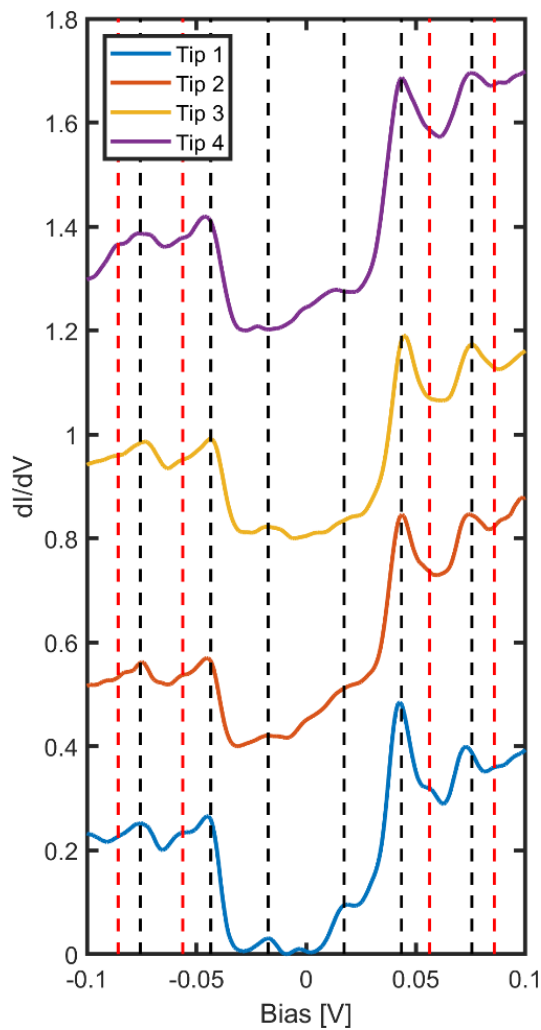

**Figure S11. Differential conductance (d$I$/d$V$) spectra (set point $V_b$ = −250 mV, $I_t$ = 0.5 nA) at DCA molecular centre within DCA$_3$Cu$_2$/Ag(111), measured with various tips.** We observe peaks at ±17.3 mV, ±43.5 mV and ±75.6 mV (black dashed lines; reported in main text), and subtler features at ±56.3 mV, ±85.8 mV (red dashed lines; not reported in main text).

For the Cu$_A$ and DCA anthracene extremity sites, our fitting procedure yielded, at 4.4 K, off-Fermi satellite Lorentzian peaks with energy positions ~±43 and ±73 mV, and a width $\Gamma \approx$ 15 mV (Figs. 3a-c of main text). This lower number of significantly broader off-Fermi satellite peaks (in comparison to the DCA centre with five satellite Lorentzian peaks and with $\Gamma \approx$ 7 mV) could be explained by the complexity of the near-Fermi spectral structure with the strong Kondo ZBP, hindering an accurate fit of all off-Fermi satellite features.

Given the symmetry of all these features with respect to the Fermi level ($V_b$ = 0), we performed DFT (with $U$ = 0) calculations of the low-energy (< 100 meV) vibrational modes for a single DCA molecule with each of its two cyano groups coordinated to a Cu atom, in the gas phase. In particular, we focus on vibrational modes with a symmetry allowing to be excited by tunneling electrons. To determine these symmetry-allowed vibrational modes, we calculated the displacement of all atoms of a negatively charged [Cu$_2$DCA]$^{-1}$ complex relative to the atoms' positions in a neutral Cu$_2$DCA complex. The vibrational modes were calculated as eigenvalues of the Hessian matrix of energy with respect to atomic position, which has eigenvectors of vibrational displacement multiplied by the square root of the atomic mass[12,13]. We define an overlap as the dot product between these vibrational eigenvectors, and the (normalized) vector resulting from the product of the displacement vector from charging and the square root of the atomic masses. The vibrational modes associated with a non-zero overlap are those whose symmetry allow for excitation by tunneling electrons. Of all the calculated vibrational modes, only four had a non-zero overlap (Fig. S12), with eigenenergies similar to those of the observed d$I$/d$V$ satellite peaks at ±17.3 mV, ±43.5 mV, ±56.3 mV, ±85.8 mV. We note, however, that this method does not explain the peak observed at ±75.6 mV. Our calculations were performed for a flat gas-phase molecular complex; on Ag(111), the Cu-coordinated DCA may undergo some conformational changes. In particular, note that the two different Cu atoms within the MOF on Ag(111), Cu$_A$ and Cu$_B$ (Fig. 1 of main text), with slightly different adsorption heights, might induce a slight tilt of DCA . This could arguably result in other vibrational modes (e.g., out-of-plane, with eigenenergy close to that associated with the d$I$/d$V$ peak at ±75.6 mV) with a symmetry enabling inelastic excitation by tunneling electrons.

| 17.3 meV | 43.1 meV | 58.4 meV | 83.4 meV |

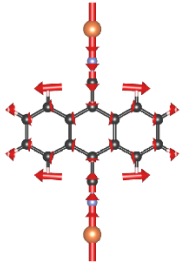 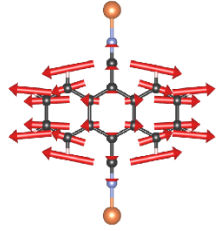 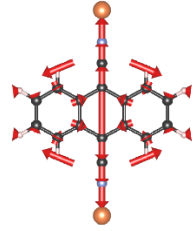 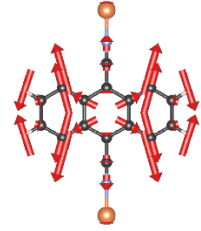

**Figure S12. DFT-calculated vibrational modes of a single Cu$_2$DCA molecular complex in the gas phase**. The length of the red arrows represents the magnitude and direction of the atoms' displacements. These vibrational modes have a symmetry that allows for inelastic excitation by tunnelling electrons, with eigenenergies similar to that of experimentally observed off-Fermi d$I$/d$V$ peaks at the DCA centre.

Typically, these inelastically excited vibrational modes appear as step-like d$I$/d$V$ features, near and symmetric in energy with respect to the Fermi level. Here, we attribute the aforementioned off-Fermi, energy-symmetric d$I$/d$V$ peaks to phonon-assisted Kondo tunneling[11,14], resulting from coupling between local magnetic moments and molecular vibrational modes. The coexistence of d$I$/d$V$ peaks and steps at ~±43 mV implies that the rates for phonon-assisted Kondo tunneling and for inelastic excitation of molecular vibrational modes at this energy are comparable. Note that neither the zero-bias Kondo resonance nor the step at ~±43 mV were observed for DCA$_3$Cu$_2$ on Cu(111)[1] or on graphene (G)/Ir(111)[15]. In the case of DCA$_3$Cu$_2$/Cu(111), a stronger interaction and hybridization between the kagome system and the more reactive Cu(111) substrate could hinder the presence of localized magnetic moments and therefore of Kondo effect. In our case on Ag(111), the increased Fermi-level LDOS given by the Kondo resonance could enhance step-like features due to inelastic tunneling[16], further explaining the absence of d$I$/d$V$ step-like features related to inelastically excited molecular vibrations for DCA$_3$Cu$_2$/Cu(111). In the case of DCA$_3$Cu$_2$/G/Ir(111), the absence of a Kondo resonance and of inelastic tunnelling features could be due to the small Fermi-level DOS at the graphene Dirac point, in comparison to Ag(111).

Similar to the ZBP, these off-Fermi satellite peaks (for all Cu, DCA centre and anthracene extremity sites) decay in magnitude and broaden as the temperature increases (Figs. 3a-c of main text). However, a quantitative temperature dependence of these off-Fermi features could not be established within a significant temperature range, also due to the intricacy of the near-Fermi d$I$/d$V$ spectra at non-zero bias voltages. It important to note that this does not affect the quantitative temperature-dependent description of the ZBP.

## S7. Band structure calculations: DFT+$U$

Figure S13a shows the DFT+$U$ calculated band structure for the freestanding DCA$_3$Cu$_2$ kagome metal-organic framework (MOF) with $U$ = 0 eV and a charge depletion of 0.25 electrons per primitive unit cell (UC), that is, p-doped, with the Fermi level below the Dirac point. This charge depletion was considered in order to match the MOF-to-surface electron transfer calculated for DCA$_3$Cu$_2$ kagome structure on Ag(111) (see below). The kagome valence band (VB) structure hosts two Dirac bands, and a flat band at ~0.2 eV with respect to the Dirac point. The projected density of states onto the DCA LUMO (blue) and Cu 3$d$ (green) states show that these kagome bands have a largely molecular character with little (but non-zero) contribution from the Cu 3$d$ states. Note that here, for $U$ = 0 eV, these bands are not spin-polarized (that is, there is no energy difference between spin 'up' and spin 'down' states). Figure S13b shows the DFT+$U$ calculated band structure for the freestanding DCA$_3$Cu$_2$ kagome MOF with $U$ = 3eV (and a charge depletion of 0.25 electrons per UC), including spin polarization. These bands resemble that of the case with $U$ = 0 eV (Fig. S13a), with some energy splitting between the spin 'up' (magenta) and spin 'down' (cyan) bands. For $U$ > 0 eV, there is a break in symmetry in the electronic band structure due to a break in real-space symmetry (due to an asymmetric spin density configuration), with Brillouin zone (BZ) boundary points $K_1$ and $K_2$, as well as $M_1$, $M_2$ and $M_3$ (insets of Figs. S13a-b), becoming inequivalent. However, the band structure along high symmetry reciprocal space directions (*e.g.*, $\Gamma - M_1 - K_1 - \Gamma$ compared to $\Gamma - M_2 - K_2 - \Gamma$, shown in Fig. S13b) remains qualitatively similar. Figures S13c-d show the DFT+$U$ calculated band structure for the DCA$_3$Cu$_2$ kagome MOF on Ag(111) with $U$ = 0 eV (reproduced here from the main text Fig. 4a) and $U$ = 3 eV, respectively. For $U$ = 3 eV (Fig. S13d), we observe that the bands are spin polarized (magenta and cyan circles are projections onto MOF spin 'up' and spin 'down' orbitals, respectively), similarly to the freestanding case (Fig. S13b).

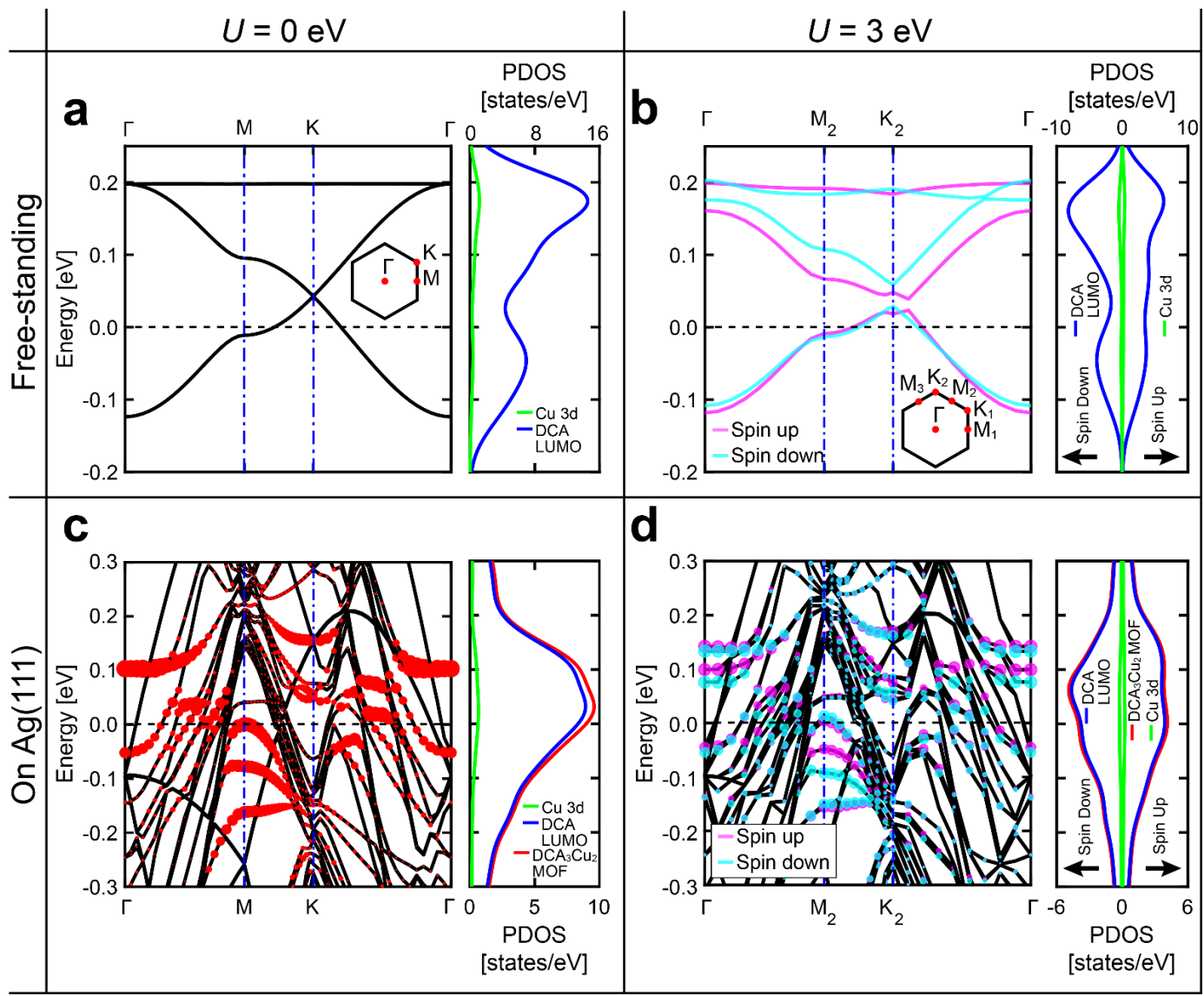

**Figure S13. DFT+$U$ calculated valence band structure for freestanding DCA$_3$Cu$_2$ and DCA$_3$Cu$_2$/Ag(111), for $U$ = 0 and 3 eV. a-b**, Band structure of freestanding DCA$_3$Cu$_2$ (depleted by 0.25 electrons per UC, that is, p-doped) with $U$ = 0 (a) and 3 eV (b). Insets: Brillouin zone with high-symmetry points, with symmetry breaking in (b) due to the increased electron-electron interactions. **c-d**, Band structure for DCA$_3$Cu$_2$ on Ag(111) with $U$ = 0 (c) and 3 eV (d). Density of states projected (PDOS) onto MOF orbitals are shown to the right. All density of states were calculated with a Gaussian broadening of 0.05 eV. In (c), red circles are projections onto DCA$_3$Cu$_2$ orbitals. In (b) and (d) the band structure and PDOS are spin-polarized [magenta and cyan curves: spin 'up and spin 'down'; magenta (cyan) circles in (d): projection onto DCA$_3$Cu$_2$ spin 'up' (spin 'down', respectively) orbitals].

For comparison, Figure S14 shows the DFT+$U$ calculated band structure for the DCA$_3$Cu$_2$ kagome MOF on Ag(111) and Cu(111) (with $U$ = 0 eV). It is important to note that, while both systems show a qualitatively similar degree of interaction between MOF and noble metal surface, with a significant perturbation of the Dirac and flat bands of the freestanding kagome electronic structure (compare with Fig. 13a), the DCA$_3$Cu$_2$/Ag(111) system exhibits remnants of the flat band near the Γ-point, not present on Cu(111). Moreover, the MOF-to-surface electron transfer (that is, degree of p-doping) is significantly larger on Cu(111) than on Ag(111) (see below). This hints towards an interaction of the DCA$_3$Cu$_2$ MOF with Cu(111) stronger than with Ag(111), consistent with the higher chemical reactivity[17] of Cu.

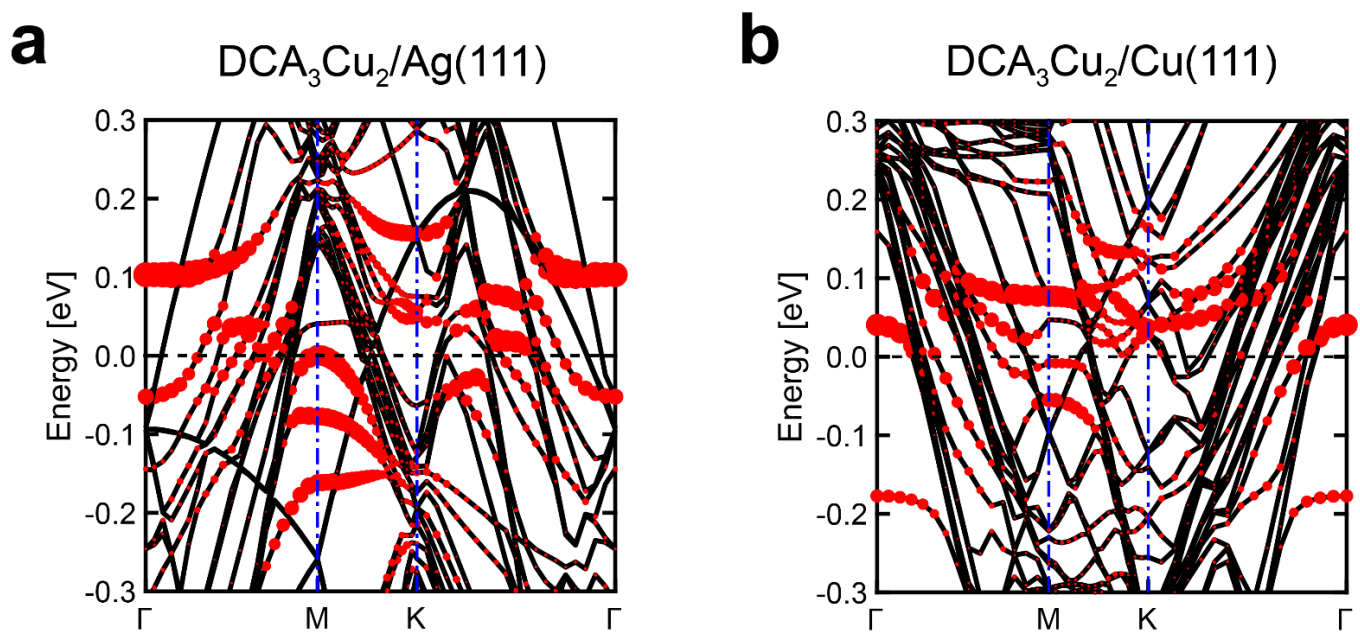

**Figure S14. DFT calculated band structure of DCA$_3$Cu$_2$/Ag(111) and DCA$_3$Cu$_2$/Cu(111), with $U$ = 0 eV. a,** DCA$_3$Cu$_2$ kagome on Ag(111). **b,** DCA$_3$Cu$_2$ kagome on Cu(111). Red circles: projections onto DCA$_3$Cu$_2$ orbitals.

## S8. MOF-to-substrate electron transfer

We performed DFT (with $U$ = 0) calculations of the band structure of kagome $DCA_3Cu_2$ on Ag(111) and Cu(111), as a function of MOF-to-surface distance (Fig. S15). We considered the (electronically and structurally) relaxed on-substrate geometry (corresponding to a distance of 0 Å in Fig. S15), translated the $DCA_3Cu_2$ kagome MOF vertically away from the noble metal surface, and then performed a single point electronic calculation, relaxing the electronic degrees of freedom to acquire the charge density and resulting band structure. As the MOF-surface distance increases (> ~2 Å), the band structure regains its freestanding character, with its two Dirac bands and one flat band, with the Fermi level below the Dirac points [for both Ag(111) and Cu(111)]. That is, $DCA_3Cu_2$ kagome MOF tends to transfer electrons to the surface, becoming p-doped. Note that increasing $U$ in the DFT+$U$ calculations does not significantly affect this number of electrons transferred (see following section).

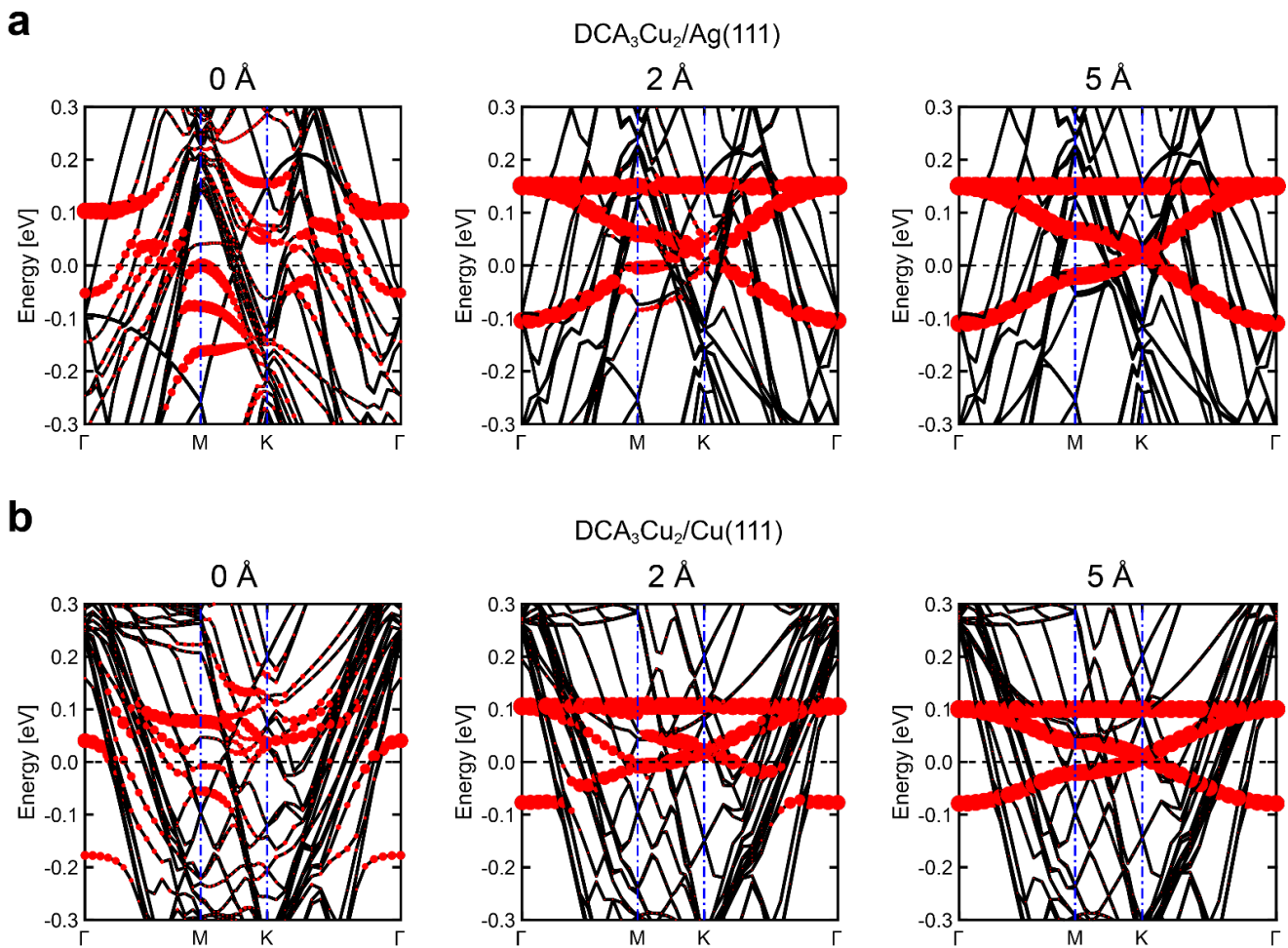

**Figure S15. DFT ($U$ = 0) calculated band structures for DCA$_3$Cu$_2$ on Ag(111) and Cu(111) as a function of MOF-surface (vertical) distance. a,** DCA$_3$Cu$_2$ kagome on Ag(111). **b**, DCA$_3$Cu$_2$ kagome on Cu(111). A displacement of 0 Å corresponds to the relaxed on-surface structure. Larger distances correspond to the relaxed MOF being vertically translated away from the surface. Red circles: contributions from MOF orbitals to the overall band structure.

Based on this observation, we performed charge analysis (Bader and DDEC; see Methods) and calculated the number of electrons transferred from MOF to surface (Fig. S16). For both Ag(111) and Cu(111), DDEC charge analysis shows that the $DCA_3Cu_2$ kagome structure tends to transfer electrons to the substrate, for equilibrium and increasing MOF-surface distances, and with a greater electron transfer on Cu(111) (~0.67 electron per UC at equilibrium distance) than on Ag(111) (~0.22 electrons per UC at equilibrium distance). This is consistent with the calculated band structures in Fig. S15. It is also consistent with the different work functions[18] of the $DCA_3Cu_2$ kagome MOF (~3.95 eV based on subtraction of the Fermi level from the electrostatic potential in vacuum), Ag(111) (4.74 eV) and Cu(111) (4.94 eV). The amount of DDEC-calculated transferred electrons decreases monotonically as the MOF-surface distance increases. At large distances (> ~1 Å), Bader and DDEC charge analysis agree. However, they disagree at short distance where there is a strong overlap of the $DCA_3Cu_2$ and substrate orbitals, and where grid-based ambiguities in determining Bader partitions can lead to substantial errors. DDEC relies on fitting charge to spherical functions and does not depend on a grid. As such, we conclude that DDEC provides a measure of electron transfer which is more accurate than Bader in these systems.

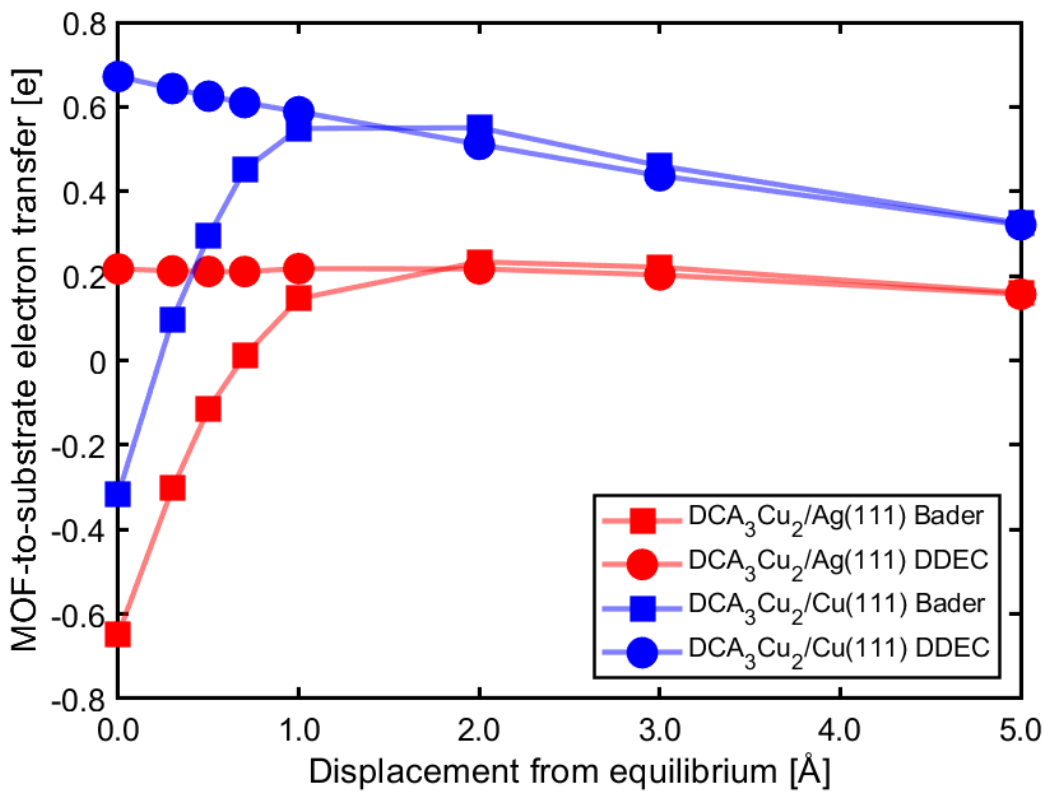

**Figure S16. MOF-to-substrate electron transfer (per primitive unit cell) resulting from DFT ($U$ = 0) calculations, and Bader and DDEC charge analysis, for Ag(111) and Cu(111), as a function MOF-surface distance.** A displacement of 0 Å corresponds to the relaxed on-surface structure. Larger distances correspond to the relaxed MOF being vertically translated away from the surface by the specified distance.

### S9. Magnetic moment and spin density calculations

DFT+$U$ local magnetic moment calculation

Figure 4c of the main text shows the average magnitude of the magnetic moment per DCA molecule within the DCA$_3$Cu$_2$/Ag(111), calculated by DFT+$U$, as a function of $U$ (where $U$ is a correction accounting for the Coulomb interaction between Cu 3$d$ electrons; see Methods in the main text). We performed similar calculations for DCA$_3$Cu$_2$ on Cu(111) (see Fig. S17a). The calculated MOF-to-surface electron transfer varies negligibly with $U$, for both Ag(111) (~0.2 electrons per MOF primitive unit cell) and Cu(111) (between 0.6 and 0.7 electrons). The calculated average local magnetic moment, $\sqrt{\langle m^2 \rangle}$, per DCA in the MOF on Cu(111) is ~0 $\mu_B$, for values up to $U$ = 5 eV, in stark contrast to the case on Ag(111). This absence of localised magnetic moments could be due to a larger MOF-to-Cu(111) electron transfer (and hence smaller electron-electron interactions) in comparison with Ag(111); this provides an explanation for the fact that a zero-bias d$I$/d$V$ resonance related to the Kondo effect has not been observed for DCA$_3$Cu$_2$/Cu(111)[1].

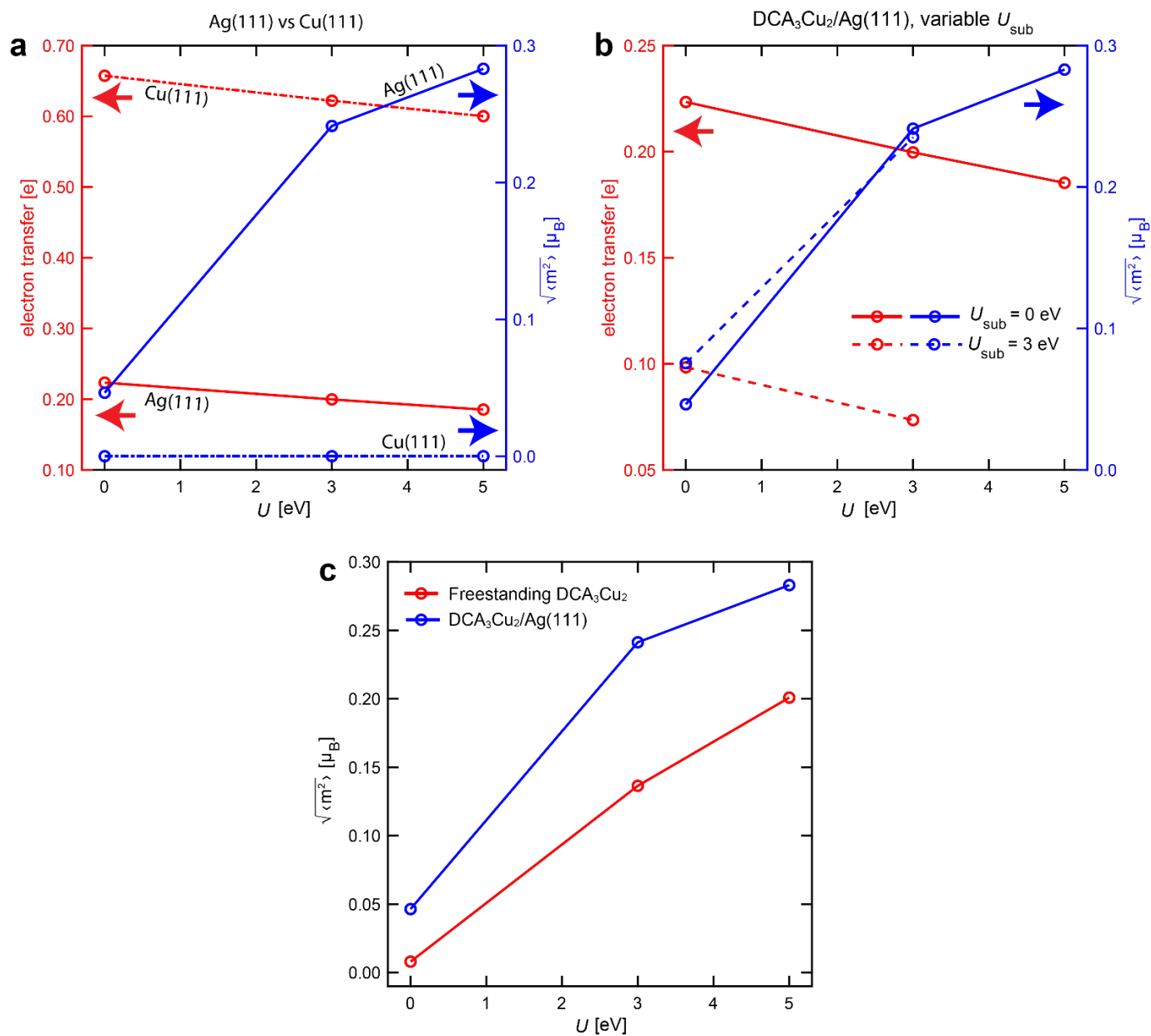

**Figure S17. Electron transfer and average local magnetic moment calculated via DFT+$U$ for freestanding DCA$_3$Cu$_2$, DCA$_3$Cu$_2$/Ag(111) and DCA$_3$Cu$_2$/Cu(111). a**, Average local magnetic moment per DCA molecule (blue, right axis) and DDEC MOF-to-substrate electron transfer per MOF primitive unit cell (red, left), as a function of $U$, for DCA$_3$Cu$_2$/Ag(111) (solid lines) and DCA$_3$Cu$_2$/Cu(111) (dashed). **b**, Average local magnetic moment per DCA molecule (blue, right) and DDEC MOF-to-substrate electron transfer per MOF primitive unit cell (red, left), as a function of $U$, for DCA$_3$Cu$_2$/Ag(111), for different values of $U_{sub}$ [correction term accounting for interactions between Ag(111) $d$ electrons]. **c**, Average local magnetic moment per DCA molecule as a function of $U$, for freestanding DCA$_3$Cu$_2$ (red line; p-doped

with 0.25 electron per MOF primitive unit cell removed) and $DCA_3Cu_2$/Ag(111). $U_{sub}$ = 0 throughout main text and SI, unless indicated.

It is important to note that in all our DFT+$U$ calculations (in the main text and SI), $U$ represents a correction term accounting for the Coulomb interaction between Cu 3$d$ electrons within the $DCA_3Cu_2$ MOF. That is, in these calculations, corrections to the Coulomb interaction between $d$ electrons of the Ag(111) and Cu(111) substrates were not taken into account ($U_{sub}$ = 0). To explore the possible role of such substrate electron-electron interactions on the systems' electronic structure, we performed DFT+$U$ calculations on $DCA_3Cu_2$/Ag(111) for $U$ between 0 and 5 eV (correction related to interactions between MOF Cu 3d electrons), and for $U_{sub}$ = 0 and 3 eV (correction term related to interactions between substrate Ag $d$ electrons; see Fig. S17b). We find that the MOF-to-substrate electron transfer decreases while the average magnetic moment per DCA molecule remains similar with increasing $U_{sub}$ (for a fixed $U$). With $U$ = 0 eV, the average magnetic moment per DCA molecule remains small for $U_{sub}$ = 3 eV ($\sqrt{\langle m^2 \rangle} < 0.08\ \mu_B$). That is, the greatest contribution to the increase in the average local magnetic moment per DCA is provided by electron-electron interactions within the MOF, rather than by substrate electrons. For example, $\sqrt{\langle m^2 \rangle}$ per DCA is greater for $U$ = 3 eV and $U_{sub}$ = 0 eV, than for $U$ = 0 eV and $U_{sub}$ = 3 eV (see Fig. S17b). Because of this, our DFT+$U$ calculations in the main text neglect these substrate electron-electron interactions ($U_{sub}$ = 0 eV) and focus on interactions related to MOF electrons. Typically, in literature, a value of $U_{sub}$ = 0 eV is assumed for conducting substrates[19–21].

Figure S17c shows the average local magnetic moment per MOF DCA, calculated via DFT+$U$, for $DCA_3Cu_2$/Ag(111) and freestanding $DCA_3Cu_2$. To ensure a reliable comparison, we considered freestanding $DCA_3Cu_2$ with 0.25 electrons per MOF primitive unit cell removed [that is, p-doped, similar to what DFT+$U$ yields for relaxed $DCA_3Cu_2$/Ag(111)]. We find that the average magnetic moment per DCA is in reasonable agreement between freestanding $DCA_3Cu_2$ and $DCA_3Cu_2$/Ag(111), for $U$ = 0, 3, and 5 eV. The higher magnetic moments on Ag(111) can be explained by strain reducing the bandwidth (as seen in Fig. S15a compared with Fig. S13a) and the hopping $t$, thus increasing the effective $U_{MFH}/t$ which increases the magnetic moment (see following subsection). This shows that despite the significant difference in calculated band structure between freestanding $DCA_3Cu_2$ and $DCA_3Cu_2$/Ag(111) (Fig. S13), their magnetic properties remain quantitatively similar.

### DFT+$U$ and MFH calculated magnetic phase diagrams for freestanding $DCA_3Cu_2$

Our DFT+$U$ calculations support the emergence of local magnetic moments in $DCA_3Cu_2$/Ag(111) as the result of electron-electron Coulomb interactions within the MOF. These local magnetic moments are a requirement for the observed Kondo effect. As emphasised in the main text, it is important to note that these DFT+$U$ calculations do not formally capture the many-body Kondo screening of these local magnetic moments by the Ag(111) conduction electrons. The lack of Kondo screening in the DFT+$U$ calculations can allow for spin-spin interactions, which could potentially result in magnetic order of these local magnetic moments. Experimentally, such magnetic order is quenched by the Kondo screening (see section S11).

In the following, we explore magnetic order and magnetic phases in freestanding kagome $DCA_3Cu_2$ given by our DFT+$U$ and MFH calculations. *Via* DFT+$U$, we calculated the magnetic moment, $m$, for each DCA kagome site, as a function of kagome VB electron filling and $U$. An electron filling of zero corresponds to the Fermi level aligned with the bottom of the Dirac bands (that is, empty kagome VB); an electron filling of 1/3 corresponds to the charge neutrality point (neutral $DCA_3Cu_2$) with the Fermi level at the Dirac point; an electron filling of ½ corresponds to one unpaired electron per DCA kagome site (*i.e.*, electron filling is $N_e/2N_{sites}$, where $N_e$ and $N_{sites}$ are the electron number and number of kagome sites in a cell, respectively). We inspected the spatial dependence of the magnetic moment (Fig. S18), identified several distinct magnetic phases, and found a correlation between the magnetic phase and the net magnetization of the system, $\langle m \rangle$, the average local magnetic moment, $\sqrt{\langle m^2 \rangle}$, and the normalized standard deviation of $m$, $\sigma_m = \sqrt{\langle m^2 \rangle - \langle m \rangle^2}/\sqrt{\langle m^2 \rangle}$. A paramagnetic phase (PM) has a near-zero net magnetization, $\langle m \rangle$. A ferromagnetic phase (FM) has a non-zero $\langle m \rangle$ and near-zero $\sigma_m$. We found numerous spin density wave (SDW) phases with antiferromagnetic interactions; they had $\sigma_m$ close to one. We also found a pinned phase (PIN) [that is, localised spin 'down' ('up') electrons pinned to a sea of surrounding spin 'up' ('down', respectively) electrons][22], which has an intermediate $\sigma_m$ between 0 and 1, although this phase can only be resolved in a supercell.

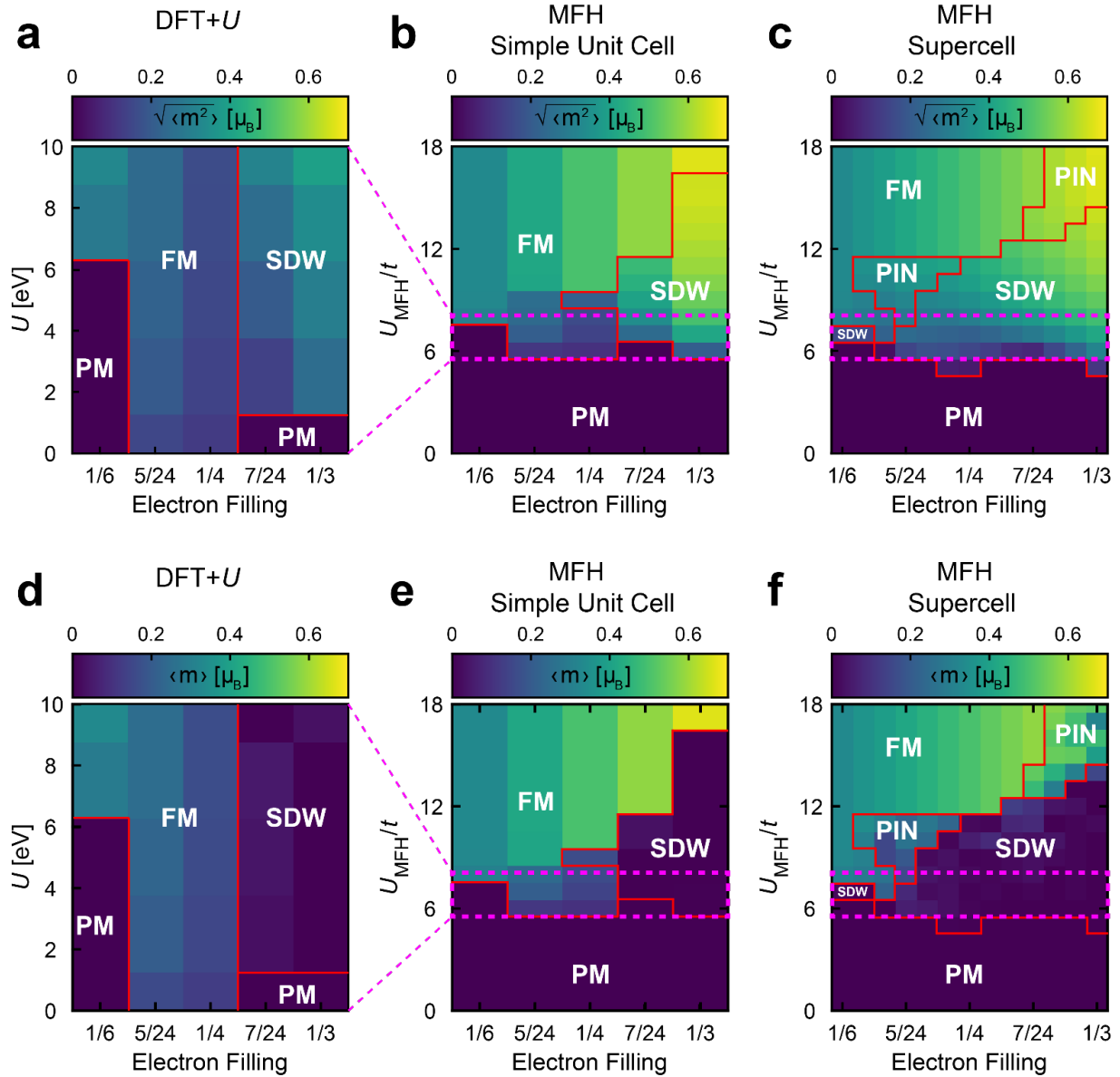

**Figure S18. Magnetic phase diagrams calculated *via* DFT+*U* and MFH model for freestanding DCA$_3$Cu$_2$ kagome structure. a**, Average local magnetic moment, $\sqrt{\langle m^2 \rangle}$, calculated by DFT+*U*, as a function of electron filling and *U*. **b**, Average local magnetic moment, $\sqrt{\langle m^2 \rangle}$, calculated by MFH model (simple primitive unit cell consisting of three kagome sites), as a function of electron filling and $U_{MFH}$. **c**, Average local magnetic moment, $\sqrt{\langle m^2 \rangle}$, calculated by MFH model (supercell consisting of 6x6 simple primitive unit cells, that is, 108 kagome sites), as a function of electron filling and $U_{MFH}$. **d-f,** Corresponding net magnetisation, $\langle m \rangle$, as calculated by DFT+*U* (d), by MFH model for a simple unit cell (e), and by MFH model for a supercell (f). Red lines delineate boundaries between different magnetic phases. PM:

paramagnetic phase; FM: ferromagnetic; SDW: spin density wave; PIN: pinned metallic. MFH model $U_{\text{MFH}}/t$ range indicated by dashed magenta regions in (b) and (c) [in (e) and (f)] corresponds to DFT+$U$ calculations $U$ range in (a) [in (d), respectively].

Figure S18a, d show the magnetic phases for freestanding DCA$_3$Cu$_2$ as a function of $U$ and of electron filling. For low electron filling (*e.g.*, ~1/6), the system remains in a paramagnetic phase for $U$ < 6 eV. With increasing electron filling and moderate $U$ (*e.g.*, $U$ = 3 eV), the system hosts non-zero, ferromagnetically ordered magnetic moments. With an electron filling of 7/24, which is close to the electron filling of DCA$_3$Cu$_2$ on Ag(111) according to our DFT+$U$ calculations, we find that the system is paramagnetic for $U$ < 1eV and is in a SDW phase for $U$ > 1 eV, consistent with DFT+$U$ results for DCA$_3$Cu$_2$ on Ag(111) (see Fig. 4c of main text).

We further explored potential magnetic phases of freestanding kagome DCA$_3$Cu$_2$ based on our MFH model (see Methods in main text). We tested our MFH model for electron fillings between ⅙ and ⅓, and for an on-site interaction, $U_{\text{MFH}}$, between 0 and 18$t$. For a system where the unit cell consists of three kagome sites ('simple unit cell'), the average local magnetic moment per kagome site generally increases with increasing $U_{\text{MFH}}$ (due to the energy penalty of spin 'up' and 'down' electrons occupying the same kagome lattice site) and increasing electron filling (due to more electrons interacting with each other), as shown in Figs. S18b, e. For an electron filling of 7/24 [corresponding to kagome DCA$_3$Cu$_2$ depleted of 0.25 electrons per MOF primitive unit cell, close to the MOF-to-surface electron transfer suggested by DFT+$U$ for DCA$_3$Cu$_2$ on Ag(111)] and intermediate $U_{\text{MFH}}$, a SDW phase is energetically favourable. Ferromagnetism occurs at higher $U_{\text{MFH}}$, where all electrons having the same spin is energetically favourable to electrons with opposite spin which can interact. The onset of FM occurs at a lower $U_{\text{MFH}}$ for lower electron fillings, which can be understood from Stoner ferromagnetism – a smaller number of electrons means

S40

depopulating one spin channel to make a FM configuration has a smaller increase in band energy, which makes an FM configuration more energetically favourable.

To extend our 'simple unit cell' MFH calculations above, we also explored different magnetic phases given by our MFH model for a 'supercell' kagome structure, where the unit cell consists of 6x6 'simple unit cells' (that is, 3 x 6 x 6 = 108 kagome sites; see Methods). The resulting magnetic phase diagram (Figs. S18c, f) is in qualitative agreement with the results for the simple primitive unit cell calculation (Figs. S18b, e). The most notable difference is that the average local magnetic moment per kagome site, $\sqrt{\langle m^2 \rangle}$, is higher in the supercell case than in the simple unit cell case (for the same electron filling and $U_{MFH}$ values). There is also a greater preference for SDW phases, and between the FM and SDW phases there is a pinned phase. This can be explained by the limited number of magnetic configurations in the simple cell case in comparison to the supercell, where magnetic order/disorder can be expressed more freely. Figure S19 shows local magnetic moments, $m$, given by our MFH model for the kagome supercell, for different $U_{MFH}$ and electron filling values. A thorough investigation of magnetic order in this system is beyond the scope of this work; magnetic order in a kagome lattice at 1/3 filling has been investigated theoretically previously[22].

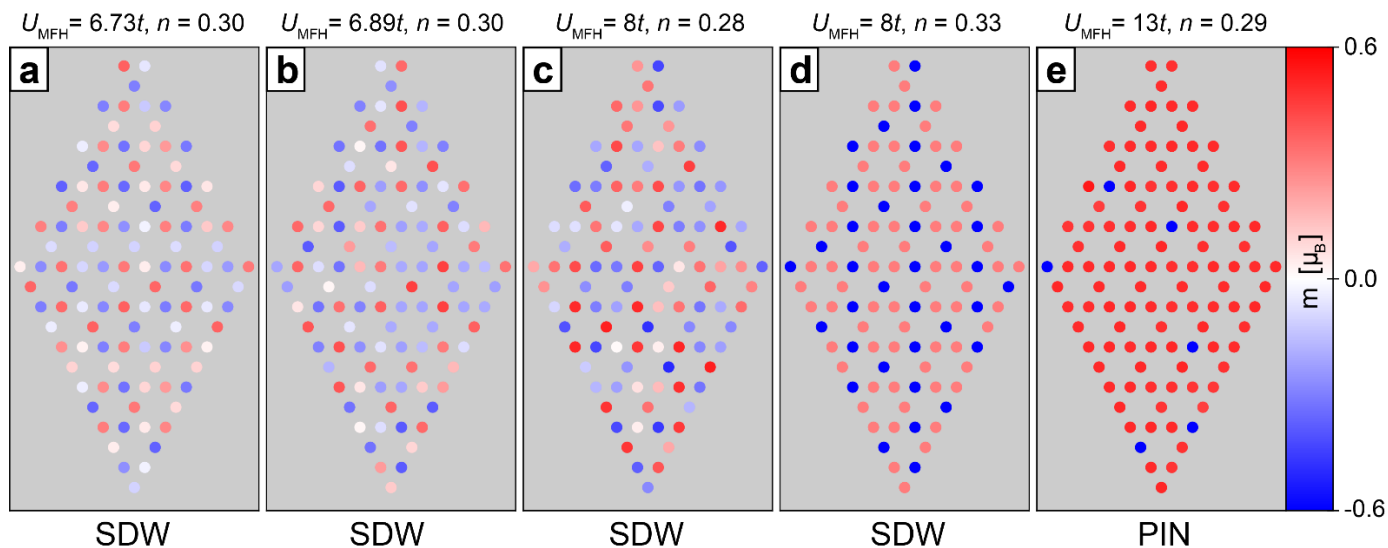

**Figure S19. Local magnetic moments, $m$, at kagome sites calculated by MFH model for selected values of $U_{MFH}$ and electron filling $n$, for supercell system. a**, $U_{MFH} = 6.73t$, $n = 0.30$ determined to match local magnetic moments given by DFT+$U$ for DCA$_3$Cu$_2$/Ag(111) with $U = 3$ eV. **b**, $U_{MFH} = 6.89t$, $n = 0.30$ determined to match local magnetic moments given by DFT+$U$ for DCA$_3$Cu$_2$/Ag(111) with $U = 5$ eV. **c-e**, Selected values of $U_{MFH}$ and $n$ within SDW phase and PIN phase (indicated at bottom of plots). Note that, due to the stochastic methods used, there may exist additional spin configurations of similar energy, although qualitative features should be consistent. $n = 0$ corresponds to empty kagome valence band (Fermi level at bottom of Dirac bands); $n = 1/3$ corresponds to charge neutrality point (Fermi level at Dirac point).

Our DFT+$U$ calculations allow us to quantify parameters $t$ and $U_{MFH}$ of our MFH model. For instance, the hopping constant $t$ can be fixed based on the DFT valence bandwidth of freestanding DCA$_3$Cu$_2$ with $U = 0$, that is, such that $6t \approx 321$ meV, hence $t \approx 53.5$ meV. The on-site Coulomb energy $U_{MFH}$ can then be estimated such that the local magnetic moments, $m$, given by the MFH model at the three kagome sites of the simple primitive unit cell match those given by DFT+$U$ for the three DCA molecules within the primitive unit cell of the DCA$_3$Cu$_2$ kagome MOF on Ag(111), using the electron filling as calculated by DDEC. For $U = 3$ eV (5 eV), we found similar MFH local magnetic moments for $U_{MFH} = 6.73t = 0.360$ eV ($U_{MFH} = 6.89t = 0.369$ eV, respectively). We estimate that $U$ = 0 eV (negligible magnetization given by DFT+$U$) corresponds to the MFH model case with $U_{MFH}$ < 6.5$t$.

We also determined the MFH model $U_{MFH}$ parameter based on DFT+$U$ calculations for freestanding DCA$_3$Cu$_2$ with an electron filling of 7/24 (corresponding to the kagome valence band structure being depleted by 0.25 electrons per MOF primitive unit cell). For $U$ = 3 eV in the DFT+$U$ calculations, we found similar local magnetic moments given by the MFH model with $U_{MFH}$ = 6.49$t$ = 0.347 eV. For $U$ = 5 eV, $U_{MFH}$ = 6.72$t$ = 0.360 eV. For an electron filling between 1/6 and 1/3, and for $U$ in our DFT+$U$ calculations up to 10 eV (Fig. S18a), the MFH model resulted in a similar overall magnetization and similar local magnetic moments for $U_{MFH}$ between 5.5$t$ (0.29 eV) and 8.5$t$ (0.45 eV). Notably, the magnetic phase diagrams given by DFT+$U$ (Figs. S18a, d) for 0 < $U$ < 10 eV, and by the (simple primitive unit cell) MFH model (Figs. S18b, e) for 5.5$t$ (0.29 eV) < $U_{MFH}$ < 8.5$t$ (0.45 eV), are in very good qualitative agreement. Combined with the agreement in freestanding DCA$_3$Cu$_2$ band structure calculated by the MFH model and by DFT+$U$ calculations, this corroborates the ability of our MFH model to reproduce the DFT results and to rationalise the electronic and magnetic properties of the DCA$_3$Cu$_2$ kagome system.

Previous theoretical work[23] on the freestanding DCA$_3$Cu$_2$ kagome system estimated an effective characteristic energy on the order of ~1 eV for Coulomb interactions between electrons on DCA kagome sites, significantly larger than that ($U_{MFH}$ ≈ 400 meV) estimated for our MFH model from DFT+$U$ calculations. This discrepancy can be explained by the fact that DFT typically over-delocalises electrons, which could lead to an underestimate of electron interactions (especially in the carbon $\pi$ orbitals where our $U$ correction does not apply), and that even DFT+$U$ using approximate functionals does not properly describe strong many-body effects[13,24]. That is, correlations between DCA electrons located at kagome sites could possibly be stronger than estimated in this work. Our study provides a lower limit for these interactions, with characteristic interaction energies that are already significantly larger than the kagome valence bandwidth.

## S10. Effect of structural disorder in DCA$_3$Cu$_2$ on Ag(111): MFH calculations

Our experimental results show a substantial degree of disorder of the DCA$_3$Cu$_2$ kagome structure on Ag(111); Fig. 1 of main text. Here, using our MFH model (see Methods in main text), we explore the effect of such disorder on the magnetic properties of DCA$_3$Cu$_2$/Ag(111). We modelled disorder by applying random variations to the nearest-neighbour hopping parameters, $t_{ij}$, and on-site energies, $\varepsilon_i$, within the supercell (that is, 6x6 simple unit cells, where each simple unit cell includes three kagome sites). Specifically, the Hamiltonian for this disordered MFH model becomes:

$$H = -\sum_{\langle i,j \rangle, \sigma} \left( t_{ij} c_{i,\sigma}^\dagger c_{j,\sigma} \right) + \sum_{i,\sigma} \varepsilon_i n_{i,\sigma} + U_{MFH} \sum_i \left( n_{i,\uparrow} \langle n_{i,\downarrow} \rangle + n_{i,\downarrow} \langle n_{i,\uparrow} \rangle - \langle n_{i,\uparrow} \rangle \langle n_{i,\downarrow} \rangle \right)$$

where $t_{ij}$ and $\varepsilon_i$ are randomly and uniformly distributed within the intervals [$t$–$\Delta t$/2, $t$+$\Delta t$/2] and [–$\Delta \varepsilon$/2, +$\Delta \varepsilon$/2], respectively. Figures S20a, b show $\sqrt{\langle m^2 \rangle}$ as a function of $\Delta t$ and $\Delta \varepsilon$, with $U_{MFH}$ = 6.6$t$ and an electron filling of ~0.28 [close to the parameters for DFT+$U$ for DCA$_3$Cu$_2$/Ag(111) with $U$ = 3 eV; see above]. For the perfectly crystalline case, without disorder ($\Delta t$ = $\Delta \varepsilon$ = 0), $\sqrt{\langle m^2 \rangle} \cong$ 0.22 $\mu_B$, while significant disorder (e.g., $\Delta t$ = 0.4$t$ and $\Delta \varepsilon$ = 2$t$) yields $\sqrt{\langle m^2 \rangle} \cong$ 0.34 $\mu_B$. For comparison, increasing $U_{MFH}$ to 8$t$ results in, for the perfectly crystalline case without disorder, $\sqrt{\langle m^2 \rangle} \cong$ 0.37 $\mu_B$. That is, very severe disorder has an effect similar to a modest change in $U_{MFH}$ or electron filling. Also, note that the degree of disorder observed experimentally for DCA$_3$Cu$_2$/Ag(111) [where the lattice constant of DCA$_3$Cu$_2$/Ag(111) varies at most by 10%; Fig. 1 of main text and Fig. S3] is, arguably, significantly smaller than that considered here for the MFH model (with $t$ and $\varepsilon$ varying on the order of $t$). From this MFH model, we conclude that the structural disorder observed experimentally for DCA$_3$Cu$_2$/Ag(111) has little or no effect on its magnetic properties, and that the emergence of local magnetic moments [and hence the differences with DCA$_3$Cu$_2$/Cu(111)] are not due to disorder. It is possible that disorder, via Anderson localisation[25] (a dynamical effect and hence not captured by our MFH model), facilitates localisation of magnetic moments and thus enhances the Kondo effect in DCA$_3$Cu$_2$/Ag(111), in contrast with the crystalline DCA$_3$Cu$_2$/Cu(111). Describing this effect is beyond the capabilities of our MFH model.

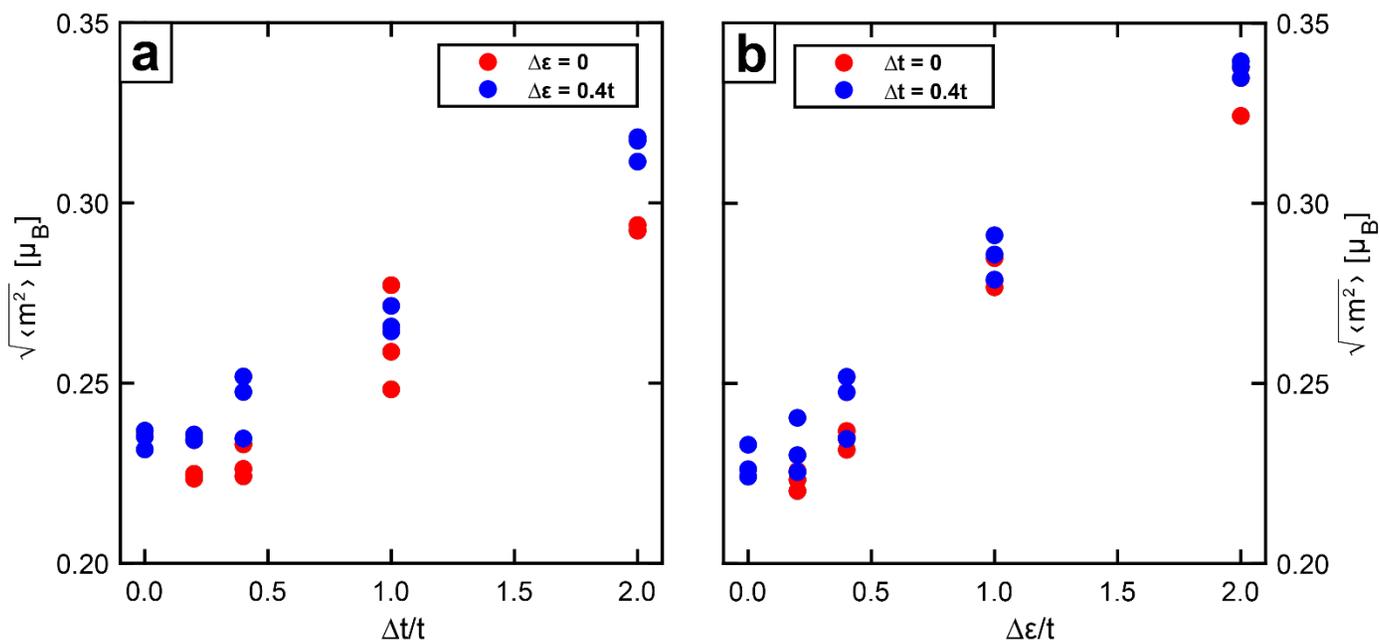

**Figure S20. Effect of structural disorder on the magnetic properties of freestanding kagome DCA$_3$Cu$_2$: MFH model (supercell). a-b**, Average local magnetic moment, $\sqrt{\langle m^2 \rangle}$, as a function of hopping disorder parameter $\Delta t$ (a; for $\Delta\varepsilon = 0$ and $0.4t$), and of on-site energy disorder parameter $\Delta\varepsilon$ (b; for $\Delta t = 0$ and $0.4t$), for $U_{MFH} = 6.6t$ and an electron filling of ~0.28. Each point represents a different configuration of random disorder.

## S11. Estimation of exchange interaction energies

### Kondo exchange interaction energy, $J_K$

Our measurements of the Kondo temperature, $T_K \sim 120$ K (see main text), allow us to estimate the Kondo exchange interaction energy, $J_K$, accounting for the coupling between the local magnetic moments within the DCA$_3$Cu$_2$ kagome structure and the conduction electrons of the underlying Ag(111) substrate. We estimate $J_K$ assuming a spin-1/2 system[26]: $k_B T_K \sim D\sqrt{2\rho J_K} e^{-1/\rho J_K}$, where $\rho$ and $D$ are the Ag(111) density of states and conduction bandwidth, respectively. With[27] $D \approx 5.5$ eV, and assuming[19,28] $\rho = 2/D$, we obtain $J_K \approx 250$ meV. This is comparable to $J_K$ reported for a single atom impurity[29].

### Direct interspin exchange interaction energy, $J_{SDW}$

We estimated the interspin exchange interaction energy, $J_{SDW}$, that would result from the direct interaction between local magnetic moments within freestanding (perfectly ordered) DCA$_3$Cu$_2$ [without Ag(111)], via our MFH model. We considered $U_{MFH}$ and an electron filling ($n$ = 0.30) that replicate the MOF-to-surface electron transfer and the local magnetic moments given by our DFT+$U$ calculations for DCA$_3$Cu$_2$/Ag(111) with $U$ = 3 and 5 eV. We then estimate $J_{SDW}$ by calculating the energy difference between the ground state magnetic configuration and the paramagnetic (that is, non-magnetic) configuration, normalized by the number of kagome sites in the system (that is, DCA molecules in DCA$_3$Cu$_2$). For $U_{MFH}$ = 6.73$t$, n = 0.30 (corresponding to $U$ = 3 eV in DFT+$U$ calculations), $J_{SDW}$ = –0.00406$t$ = –0.22 meV. For $U_{MFH}$ = 6.89$t$, n = 0.30 ($U$ = 5 eV in DFT+$U$ calculations), $J_{SDW}$ = –0.00722$t$ = –0.39 meV. These values are significantly smaller than the energy positions of the off-Fermi satellite peaks and step features observed in the near-Fermi d$I$/d$V$ spectra at the DCA molecular centre (Fig. 3 of main text). That is, these features cannot be explained as contributions to the d*I*/d*V* spectra by inelastic spin excitations (e.g., triplet-to-singlet transitions). This supports our assignment of these features to inelastic excitations of molecular vibrational modes.

## RKKY interspin interaction energy, $J_{RKKY}$

We estimated the magnitude of hypothetical Ruderman-Kittel-Kasuya-Yosida (RKKY) interactions between local magnetic moments within DCA$_3$Cu$_2$, mediated by Ag(111) conduction electrons. The RKKY exchange interaction energy, $J_{RKKY}$, is given by[30]:

$$J_{RKKY}(R) = \frac{16 J_{eff}^2 m_e k_F^4}{(2\pi)^3 \hbar^2} \left[ \frac{cos(2k_F R)}{(2k_F R)^3} - \frac{sin(2k_F R)}{(2k_F R)^4} \right]$$

where $R$ is the distance between magnetic moments, $k_F$ is the Ag (bulk) Fermi wavevector, $m_e$ is the electron mass, and $J_{eff}$ is the intra-atomic exchange parameter. One possibility to estimate the magnitude of $J_{eff}$ is to consider the Anderson impurity model[31]:

$$J_{eff} = -\frac{\Delta}{\pi N(\epsilon_0)} \frac{U}{|\epsilon_0|(U - |\epsilon_0|)}$$

where $N(\epsilon_0)$ is the Ag density of states at the energy of the impurity (estimated[32] as 3.659 atm$^{-1}$ Ry$^{-1}$), *U* is the Hubbard energy (here, associated with a localised magnetic moment, that is, with the energy cost associated with populating the localised state with two electrons), Δ is the full width at half maximum of the partially occupied localised state, and $\epsilon_0$ its energy position with respect to the Fermi level. Our d*I*/d*V* STS measurements did not allow us to clearly identify a partially occupied state below the Fermi level, despite extensive investigation (*e.g.*, see Fig. S7). For the purposes of an estimation, we assumed that *U* ~ 1 eV (motivated by our DFT+*U* calculations as well as by previous studies[23]; a larger value of U yields a lower value of $J_{eff}$). We assumed $\epsilon_0$ = 70 meV as a conservative estimation in line with previous similar experimental works[14,33], and $\Delta = 100$ meV based on previous measurements[7] of the DCA LUMO on Ag(111). We considered only interactions mediated by bulk Ag electrons (that is, bulk[27] Fermi wavevector in the equation above) since the Ag(111) Shockley surface state is completely depopulated (and therefore does not play a role here) due to scattering and confinement by the DCA$_3$Cu$_2$ kagome MOF (*i.e.*, related d*I*/d*V* features above Fermi; see SI section S2).

Under these conditions, the largest absolute value for $J_{RKKY}$ (that is, the largest peak value of $J_{RKKY}$ given by an interspin distance closest to ~0.6 nm, which is the approximate separation between DCA anthracene extremity sites) provides an upper limit for $J_{RKKY} \sim 9$ meV. This is an order of magnitude smaller than the estimated Kondo exchange interaction energy, $J_K$ (see above), and significantly smaller than the energy positions of the off-Fermi satellite peaks and step features observed in the near-Fermi d*I*/d*V* spectra at the DCA molecular centre (that we associate with molecular vibrational modes; Fig. 3 of main text and SI section S6). This small, estimated magnitude of $J_{RKKY}$ suggests that the RKKY interaction does not play a significant role in the physics of our system.